\documentclass[]{interact}

\usepackage{epstopdf}% To incorporate .eps illustrations using PDFLaTeX, etc.
\usepackage{subfigure}% Support for small, `sub' figures and tables
\usepackage{palatino}
\usepackage{wrapfig}

\usepackage[numbers,sort&compress]{natbib}% Citation support using natbib.sty
\bibpunct[, ]{[}{]}{,}{n}{,}{,}% Citation support using natbib.sty
\theoremstyle{plain}% Theorem-like structures

\theoremstyle{definition}

\theoremstyle{remark}

\usepackage{hyperref}
\usepackage[all]{hypcap}    %for going to the top of an image when a figure reference is clicked
\hypersetup{
    colorlinks,
    citecolor=blue,
    filecolor=blue,
    linkcolor=blue,
    urlcolor=blue
}
\newcommand{\fesc}{\ifmmode{f_{\rm esc}}\else{$f_{\rm esc}$}\fi}
\newcommand{\fescs}{\ifmmode{f_{\rm esc}^\star}\else{$f_{\rm esc}^\star$}\fi}
\newcommand{\kms}{\ifmmode{{\;\rm km~s^{-1}}}\else{km~s$^{-1}$}\fi}
\newcommand{\fgas}{\ifmmode{{f_{\rm gas}}}\else{$f_{\rm gas}$}\fi}
\newcommand{\cubecm}{\ifmmode{{\rm cm^{-3}}}\else{cm$^{-3}$}\fi}
\newcommand{\ztwo}{\ifmmode{{\rm [Z_2/H]}}\else{[Z$_2$/H]}\fi}
\newcommand{\zthree}{\ifmmode{{\rm [Z_3/H]}}\else{[Z$_3$/H]}\fi}
\newcommand{\lsim}{\lower0.3em\hbox{$\,\buildrel <\over\sim\,$}}
\newcommand{\gsim}{\lower0.3em\hbox{$\,\buildrel >\over\sim\,$}}

\newcommand{\sfr}{M$_\odot$ yr$^{-1}$ Mpc$^{-3}$}
\newcommand{\hsfr}{M$_\odot$ yr$^{-1}$}

\newcommand{\eavg}{\ifmmode{\langle E_\gamma \rangle}\else{$\langle E_\gamma \rangle$}\fi}

\newcommand{\Ms}{\ifmmode{\textrm{M}_\odot}\else{M$_\odot$}\fi}
\newcommand{\vrms}{\ifmmode{v_{\rm rms}}\else{$v_{\rm rms}$}\fi}

\newcommand{\hh}{H$_2$}

\newcommand{\tvir}{\ifmmode{T_{\rm{vir}}}\else{$T_{\rm{vir}}$}\fi}
\newcommand{\mvir}{\ifmmode{M_{\rm{vir}}}\else{$M_{\rm{vir}}$}\fi}
\newcommand{\rvir}{\ifmmode{r_{\rm{vir}}}\else{$r_{\rm{vir}}$}\fi}

\newcommand{\lya}{Ly$\alpha$}
\newcommand{\jj}{\ifmmode{J_{21}}\else{$J_{21}$}\fi}
\newcommand{\flw}{\ifmmode{F_{LW}}\else{$F_{LW}$}\fi}
\newcommand{\kph}{\ifmmode{k_{\rm ph}}\else{$k_{\rm ph}$}\fi}

\newcommand{\zsun}{\ifmmode{\rm\,Z_\odot}\else{$\rm\,Z_\odot$}\fi}

\newcommand{\hii}{H {\sc ii}}

\newcommand\unit[1]{\; \textrm{#1}}

\begin{document}

\articletype{INTRODUCTORY REVIEW ARTICLE}

\title{Cosmic Reionization}

\author{
  \name{John H. Wise\textsuperscript{$\ast$}\thanks{Email: jwise@gatech.edu}}
  \affil{\textsuperscript{$\ast$} Center for Relativistic Astrophysics,
    School of Physics, Georgia Institute of Technology, Atlanta, GA, USA}
}

\maketitle

\begin{abstract}
  
  The universe goes through several phase transitions during its
  formative stages.  Cosmic reionization is the last of them, where
  ultraviolet and X-ray radiation escape from the first generations of
  galaxies heating and ionizing their surroundings and subsequently
  the entire intergalactic medium.  There is strong observational
  evidence that cosmic reionization ended approximately one billion
  years after the Big Bang, but there are still uncertainties that
  will be clarified with upcoming optical, infrared, and radio
  facilities in the next decade.  This article gives an introduction
  to the theoretical and observational aspects of cosmic reionization
  and discusses their role in our understanding of early galaxy
  formation and cosmology.

\end{abstract}


\section{Historical Preface}

Astronomers use distant objects as flashlights to peer through the
vast space between galaxies, the so-called intergalactic medium (IGM).
Intervening gas clouds absorb light in particular atomic transitions,
producing an absorption spectrum, which can then be used to study the
evolution of IGM properties throughout cosmic time.  The first
quasi-stellar objects (known as QSOs or quasars) were identified in
1960 by Rudolph Minkowski\footnote{Son of Hermann Minkowski of the
  'Minkowski spacetime' in general relativity.} \citep{Minkowski60},
Allan Sandage, and Maarten Schmidt \citep{Matthews61}.  They had
star-like images, strong radio emission, and strangely placed, broad
emission lines.  Schmidt realized a few years later that the emission
lines were actually hydrogen Balmer (principle quantum number $n
\rightarrow 2$) series lines redshifted by 16\% \citep{Schmidt63}, who
concluded that an extragalactic origin was the ``most direct and least
objectable'' explanation.  This realization sparked a flurry of
research on QSOs \citep[e.g.][]{Hoyle63, Schmidt64, Sandage65,
  Osterbrock66}, which were proposed to be powered by supermassive
black holes (SMBHs) in 1964 by Edwin Salpeter \citep{Salpeter64} and
Yakov Zel'dovich \citep{Zeldovich64}, and later associated with
galactic nuclei by Donald Lynden-Bell \citep{LB69} in 1969.  These
ideas were slowly accepted, but mounting evidence, especially with
X-ray observations in the following decade
\citep[e.g.][]{Tananbaum79}, confirmed the black hole paradigm.

James Gunn and Bruce Peterson \citep{GP65} first realized in 1965 that
even if a tiny fraction ($\sim 10^{-4}$) of the IGM was neutral it
would absorb the QSO light blueward of the \lya{} wavelength at
1216~\AA{} ($n = 1 \rightarrow 2$).  However when interpreting the
most distant QSO at the time (redshifted by a factor of 2.01), there
was no such absorption, strongly suggesting that the IGM was {\it
  highly ionized} when the universe was only one-quarter of its
present age of 13.8 billion years.

Cosmic reionization is the process in which the IGM becomes ionized
and heated.  Now from many observations of the IGM and early galaxies,
we know that reionization\footnote{In this article, we refer to
  'reionization' as the reionization of hydrogen and singly-ionized
  helium.  Helium is doubly-ionized at a time 2--3 billion years after
  the Big Bang.} occurs everywhere in the IGM within a billion years
after the Big Bang.  As the first generations of galaxies fiercely
form, they provide the necessary radiation to propel this cosmological
event.  Thus, it is informative to first overview some cosmological
concepts and the astrophysics of photo-ionization in order to obtain a
fuller understanding of reionization.

\section{Background astrophysics}

\subsection{Cosmology}

Cosmology is the study of the universe in its entirety.  Large-scale
galaxy surveys suggest that the universe follows the Cosmological
Principle, stating the properties of the universe are the same for all
observers when viewed at large enough scales (greater than 300
Mpc)\footnote{1 Mpc = 3.26 million light-years = $3.086 \times 10^{24}
  \unit{cm}$}.  In other words, the large-scale structure of the
universe would be indistinguishable when traveling through space.  At
these scales, the universe is said to be {\it isotropic} and {\it
  homogeneous}.  An isotropic universe means that there is no special
direction in the universe, that is, it looks the same in all
directions.  A homogeneous universe is one with constant density and
the distribution of galaxies is the same wherever the observer looks.

An isotropic and homogeneous universe can be treated as one entity.
One can use general relativity to describe its dynamics, starting with
the FLRW\footnote{Friedmann-Lema{\^ i}tre-Robertson-Walker, all of who
  independently formulated this metric in the 1920's and 1930's.}
space-time metric,
\begin{equation}
  ds^2 = c^2 dt^2 - a^2(t) \left( \frac{dr^2}{1 - Kr^2} + r^2
    d\Omega^2 \right)
\end{equation}
which results in an equation of motion describing its expansion or
contraction.  Here $ds$, $dt$, and $d\Omega$ are intervals of
space-time, time, and angle, respectively, and $c$ is the speed of
light.  The variable $r$ is a spherical coordinate describing some
observer and is not necessarily a distance measure between two
observers.  The curvature $K$ is a constant and can be either --1, 0,
or +1, respectively corresponding to negative (elliptical), flat
(Euclidean), and positive (hyperbolic) curvature of space.

The scale factor $a(t)$ denotes the expansion of the universe, where
it is convention to take $a = 1$ at the present day.  In other words,
when the universe had only expanded to half of its current size, $a =
1/2$.  The scale factor is also related to cosmological redshift $z =
1/a - 1$, which is common measure of distance and time in cosmology.
A solution to the FLRW metric is the Friedmann equation
\begin{equation}
  \label{eqn:Hubble}
  \left( \frac{\dot{a}}{a} \right)^2 = H^2(t) = 
  \frac{8\pi G}{3} \, \rho - \frac{K c^2}{a^2}.
\end{equation}
Here $H(t)$ is the Hubble parameter that describes the expansion
(positive) or contraction (negative) rate of the universe.  Edwin
Hubble first discovered the expansion of the universe in 1927 by
determining that more distant galaxies were receding at faster
velocities.  The present day value $H_0$ is thus usually given in
units of km s$^{-1}$ Mpc$^{-1}$ that is the slope of this
relationship.

The total mass-energy density $\rho = \rho_m + \rho_r + \rho_\Lambda$
of the universe comprises
\begin{itemize}
\item non-relativistic (cold) matter ($\rho_m \propto a^{-3}$) being
  geometrically diluted as space expands,
\item radiation ($\rho_r \propto a^{-4}$) being both geometrically
  diluted and softened as its frequency is cosmologically redshifted,
  and
\item vacuum energy, the so-called cosmological constant or dark
  energy, ($\rho_\Lambda \propto a^0$) that is pervasive and uniform
  throughout the universe.
\end{itemize}
It is useful to define the critical density $\rho_{\rm c} =
3H_0^2/8\pi G$, which is obtained by solving for $\rho$ in Equation
\ref{eqn:Hubble} after setting $K=0$.  It is the dividing line between
an open universe that expands forever and a closed universe that
collapses onto itself.

Current constraints on the mass-energy components come from several
experiments and sources---cosmic microwave background, supernovae,
galaxy clusters, and large-scale structure.  They have shown that the
universe is flat ($K = 0$), and about 69\%, 26\%, and 5\% of the
mass-energy is contained in dark energy, cold dark matter (DM), and
baryons, respectively \citep{Planck18_Cosmo}.  A small fraction ($9.3
\times 10^{-5}$) is contained in radiation.  These percentages of the
$i$-th mass-energy component are always given in units of the critical
density as $\Omega_i \equiv \rho_i / \rho_{\rm c}$: $(\Omega_\Lambda,
\Omega_{\rm c}, \Omega_{\rm b}, \Omega_{\rm r}) \simeq (0.69, 0.26,
0.05, 9.3 \times 10^{-5})$.  Unless otherwise stated, we assume these
cosmological parameters in this article.

The Friedmann equation can be integrated to find that the scale factor
$a$ or $(1+z)^{-1}$ is proportional to $t^{1/2}$ in a
radiation-dominated universe, $t^{2/3}$ in a flat matter-dominated
(the so-called Einstein-de Sitter, EdS) universe, and expands
exponentially in a vacuum-dominated universe.  The latest
observational constraints suggest that the universe is flat and the
cosmological constant exists ($\rho_\Lambda > 0$), thus the scale
factor $a(t)$ is monotonically increases with time, i.e. the universe
is expanding and accelerating.

\begin{figure}[t]
  \centering
  \includegraphics[width=\textwidth]{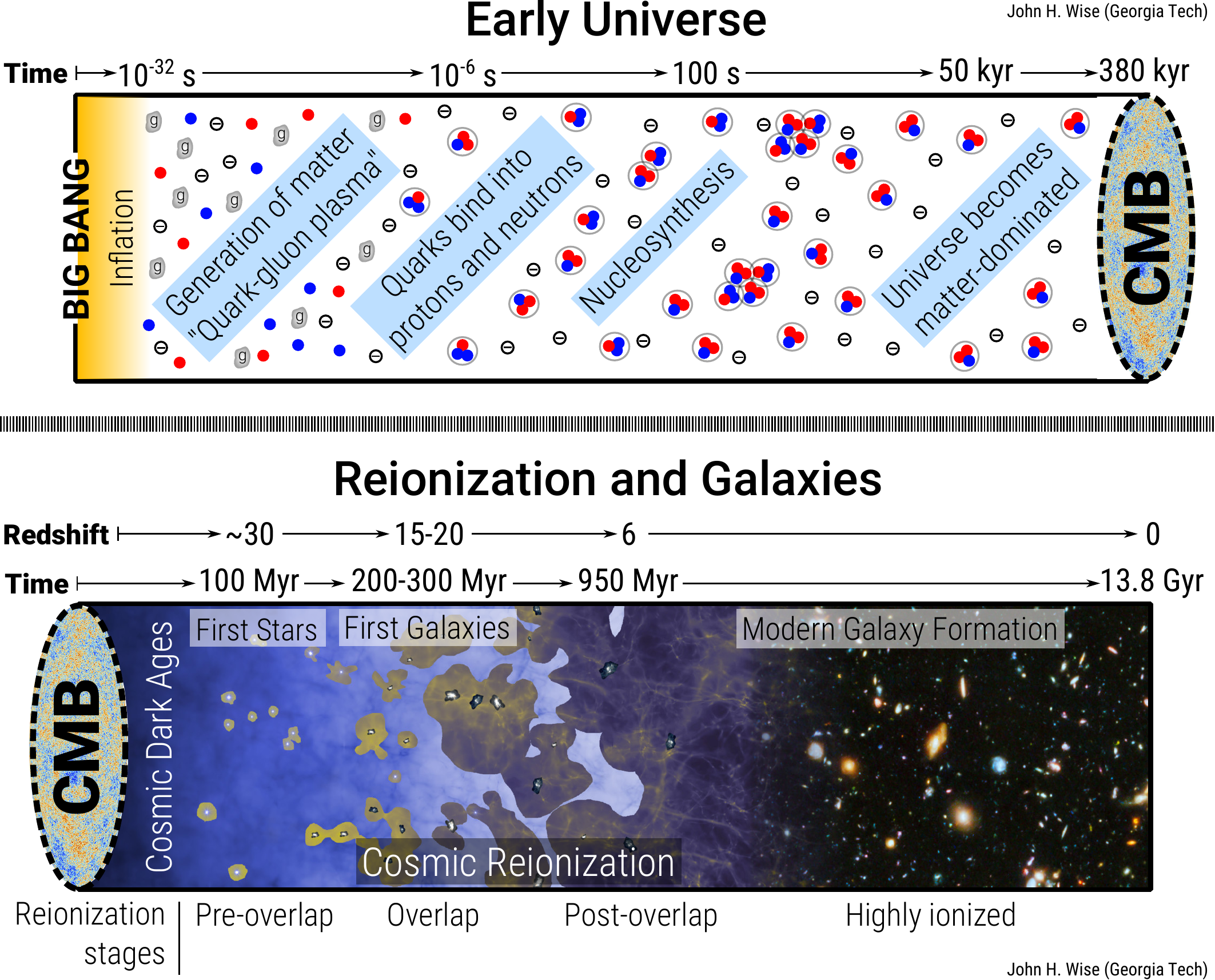}
  \caption{Cosmic timeline of the universe before (top) and after
    (bottom) recombination along with the stages of reionization.  The
    galaxy survey image is taken from the {\it Hubble Ultra Deep
      Field}.}
  \label{fig:timeline}
\end{figure}

Knowing how the universe expands as a function of time, we also know
how the mean temperature of the universe behaves, assuming it is
filled with a perfect fluid.  Starting from a hot Big Bang, the
universe undergoes several phase transitions, where some are
associated with the splitting of the four fundamental forces -- gravity,
strong, weak, and electromagnetic.  All large-scale structures are
seeded by quantum fluctuations that exponentially grew in size during
the inflationary epoch ($t \sim 10^{-36} - 10^{-32} \unit{s}$ after
the Big Bang) that eventually evolve into galaxies, shown in the
cosmic timeline in Figure \ref{fig:timeline}.

Along the way, one landmark cosmic event in the Universe occurred when
gas transitioned from an ionized to a neutral state, known as either
recombination or the surface of last scattering.  Because the density
of free electrons suddenly decreases, the photons become decoupled
from matter and stream away, creating the cosmic microwave background
(CMB).

After photons decouple, the universe at this time is a very dark and
lonely place before stars or galaxies have formed.  This epoch is
sometimes referred to as the 'Dark Ages' \citep{Rees97}.  From this
starless, neutral, and cold state, the entire universe will be
gradually reionized by nascent galaxies and their constituents.

\subsection{Large-Scale Structure}

On small scales, the universe is neither isotropic nor homogeneous.
One can look at our Milky Way and other galaxies just to see the
inhomogeneity of matter.  In the absence of light pollution, the dense
stellar fields of the Milky Way, intertwined with dark dust lanes,
stretch from horizon to horizon.  Our galaxy is a very clumpy and
active region in the universe, scattered with stellar clusters, cold
molecular clouds, warm ionized regions, and hot supernova remnants.

\begin{figure}
  \centering
  \includegraphics[width=\textwidth]{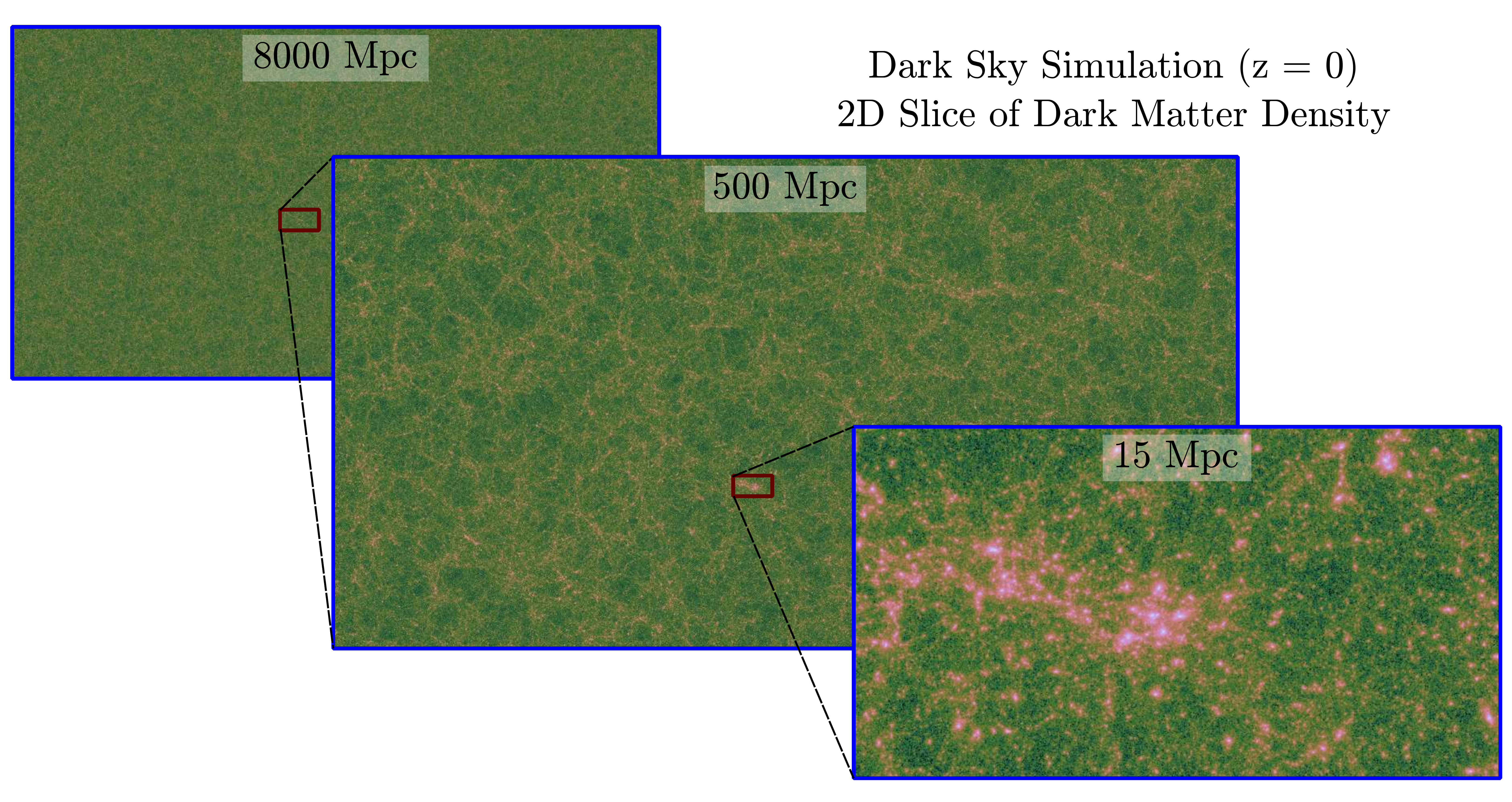}
  \caption{The Cosmic Web shown through slices of dark matter density
    in the Dark Sky Simulations \citep{Skillman14} at the present-day.
    The fields of view in the left, center, and right panels are 8000,
    500, and 15 Mpc.  The universe is homogeneous and isotropic at
    scales above 300 Mpc but below which forms large-scale structures
    (clusters, filaments, walls, and voids).}
  \label{fig:web}
\end{figure}

On slightly larger scales, these 'island universes' congregate in
groups and clusters, containing dozens and hundreds of galaxies,
respectively, which are connected to each other through long
filaments.  These cosmic connections form the cosmic web (see Figure
\ref{fig:web}) containing all cosmic structure: groups, clusters,
superclusters, filaments, walls, and voids.  When viewed in hundreds
of Mpc, galaxy number densities become uniform, and web seems to
repeat itself, suggesting the Cosmological Principle holds.

The cold dark matter (CDM) paradigm \citep{Peebles82, Blumenthal84,
  Davis85} explains cosmological large-scale structure extremely well,
predicting galaxy number densities and their clustering to great
accuracy.  Because DM constitutes most of the matter in the universe,
it controls the dynamics of large-scale structure.  In any CDM
cosmology, the building blocks are DM halos that have gravitationally
collapsed and decoupled from the expansion of the universe.  When a
halo collapses, the dark matter and gas reaches equilibrium between
the gravitational potential and its thermal and kinetic energy
\citep{LB67}.  In equilibrium, one can use a spherically symmetric
halo collapsing in an expanding universe to find that its mean density
of $18\pi^2$ times the critical density $\rho_{\rm c}$ at the collapse
time \citep{Gunn72}.  Recall that the density changes with $a^{-3} =
(1+z)^3$ from cosmological effects.  Given a halo mass $M_{\rm h}$ and
the halo mean density $\rho_{\rm h} = 18\pi^2 \rho_{\rm c} (1+z)^3$,
three halo properties can be derived.
\begin{enumerate}
\item Assuming a spherical halo, we compute the {\bf halo radius} by
  dividing the halo mass by the mean density, giving the volume, and
  we then solve for the radius,
  \begin{equation}
    r_{\rm h} = \left[ \frac{M_{\rm h}} {(4\pi/3) \rho_{\rm h} (1+z)^3} \right]^{1/3}.
\end{equation}
\item The kinematics of stars and gas within a halo can be better
  understood by relating their velocities to the value that results in
  circular motion, the {\bf circular velocity} of a system,
  \begin{equation}
    V_{\rm c} = \sqrt{ \frac{ GM_{\rm h}}{r_{\rm h}} }.
  \end{equation}
\item As the halo collapses, the gravitational potential energy can be
  converted into thermal energy and an equivalent temperature, which
  is termed the {\bf virial temperature}.  We can calculate it by
  equating the kinetic energy, $mV_{\rm c}^2/2$, of a typical gas particle
  (usually hydrogen) with mass $m$ in a circular orbit with its
  thermal energy, $k_{\rm b} T_{\rm vir}$, resulting in
  \begin{equation}
    T_{\rm vir} = \frac{mV_{\rm c}^2}{2 k_{\rm b}}
  \end{equation}
\end{enumerate}

CDM structure forms hierarchically with objects growing through a
series of halo mergers.  Galaxies exist at the centers of DM halos and
are along for the ride throughout mergers, sparking galaxy
interactions and mergers.

In this review, we are interested in the galaxies that form inside
these halos, asking ourselves the question: {\it How frequently do
  stars and galaxies form in the early universe, and generally when do
  they form?}  For a halo to host star formation, the gas must be able
to cool and condense through some radiative process.  Catastrophic
cooling and collapse occurs when a halo reaches some critical
temperature $T_{\rm vir}$ and associated mass $M_{\rm h}$.  From these
critical points, we can estimate the abundance of the first stars and
galaxies as a function of time.  Their radiation will permeate the
IGM, photoionizing and photoheating it.  To study reionization from
first principles point of view, we next overview the basic physics of
photo-ionization.

\subsection{Ionization and recombination}

\begin{figure}
  \centering
  \includegraphics[width=\textwidth]{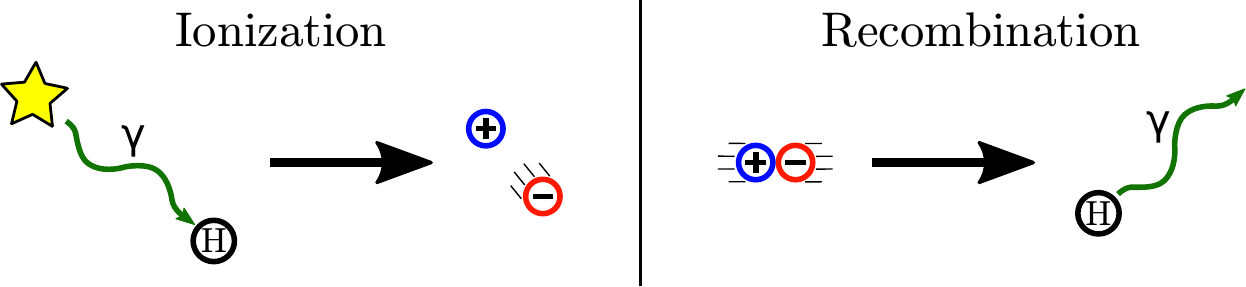}
  \caption{Cosmic reionization is governed by individual ionization
    (left) and recombination (right) events.  Photons above 13.6~eV
    will ionize neutral hydrogen, creating a free electron and proton.
    Recombination occurs more often at lower temperatures ($T <
    10^4~\mathrm{K}$) when these particles combine back into neutral
    hydrogen.  This produces a photon with 13.6~eV plus any excess
    energy from the particles' kinetic energy.}
  \label{fig:radiation}
\end{figure}

The radiation from the first luminous objects will ionize and heat the
surrounding IGM once it escapes from their host halos.  These are
known as cosmological \hii~regions.  The ionization and recombination
of hydrogen atoms ($\mathrm{H} + \gamma \leftrightarrow \textrm{H}^+ +
\textrm{e}^-$) are dominant processes in \hii~regions and are
illustrated in the Figure \ref{fig:radiation}.  Ionizations occur when
photons with energies $E > I_{\rm H} = 13.6 \unit{eV}$ interact with
neutral hydrogen atoms, where their excess energies are subsequently
thermalized.  Recombinations occur when the Coulomb force attracts
protons and electrons, which becomes efficient at temperatures $T \le
10^4 \unit{K}$.  One recombination releases a photon with an energy
that is the sum of the kinetic energy of the electron and binding
energy of the quantum state, $I_{\rm H}/n^2$.  If the electron
recombines into an excited state ($n > 1$), the electron will quickly
decay into the ground state in a series of transitions.

In a region with ionizing radiation, the gas approaches ionization
balance with the recombination rate equaling the ionization rate.
We can equate these two rates and solve for the ionization fraction
$x_e \equiv n_{\rm e}/n_{\rm H}$ of the gas at equilibrium.  Here
$n_{\rm e}$ and $n_{\rm H}$ are the electron and hydrogen number
density, respectively.

To calculate the recombination rate, we can make the following
assumptions: (i) the rate will be proportional to the product of the
number density of protons $n_{\rm p}$ and electrons $n_{\rm e}$, (ii)
the gas is pure hydrogen, giving $n_{\rm p} = n_{\rm e}$, and (iii)
the rate depends on the temperature and into which quantum state $n$
it recombines.  There are two rates associated with recombination:
\begin{description}
\item[Case A:] This rate is the sum of recombination rates into all
  electronic levels.  However if an electron recombines directly into
  the ground state, this will release a photon that can ionize another
  hydrogen atom, which is not the case if it recombines into an
  excited state.  Effectively, there are zero net recombinations in
  this case, and we can ignore the $n=1$ recombination rate.
\item[Case B:] This rate is the sum of recombination rates into all
  excited states, ignoring the ground state for reasons recently
  stated.  It is also known as the ``on-the-spot'' approximation and
  is the appropriate one to use in the ionization balance problem.
\end{description}

The ionization rate can be calculated given some ionizing radiation
flux.  Photons will be absorbed as they travel through a medium with a
neutral hydrogen number density, and the probability for a single
absorption is quantified by the photoionization cross-section
$\sigma_{\rm HI} \approx 6.8 \times 10^{-18} (E/13.6 \unit{eV})^{-3}
\unit{cm}^2$ that decreases with photon energy.  The cross-section is
zero below 13.6~eV because the photon does not have sufficient energy
to ionize hydrogen.

By equating the recombination and ionization rates, we can show that
\hii{} regions are {\it highly ionized}, having ionization fractions
nearly equal to one and neutral hydrogen fractions around $10^{-4}$.
They are thus well described by a fully ionized plasma.  The
ionization flux, however, decreases as the distance squared from the
star, resulting in the medium being the most ionized near radiation
sources with it decreases rapidly with increasing distance.

\subsection{Evolution of an \hii~region}

\begin{figure}
  \centering
  \includegraphics[width=\textwidth]{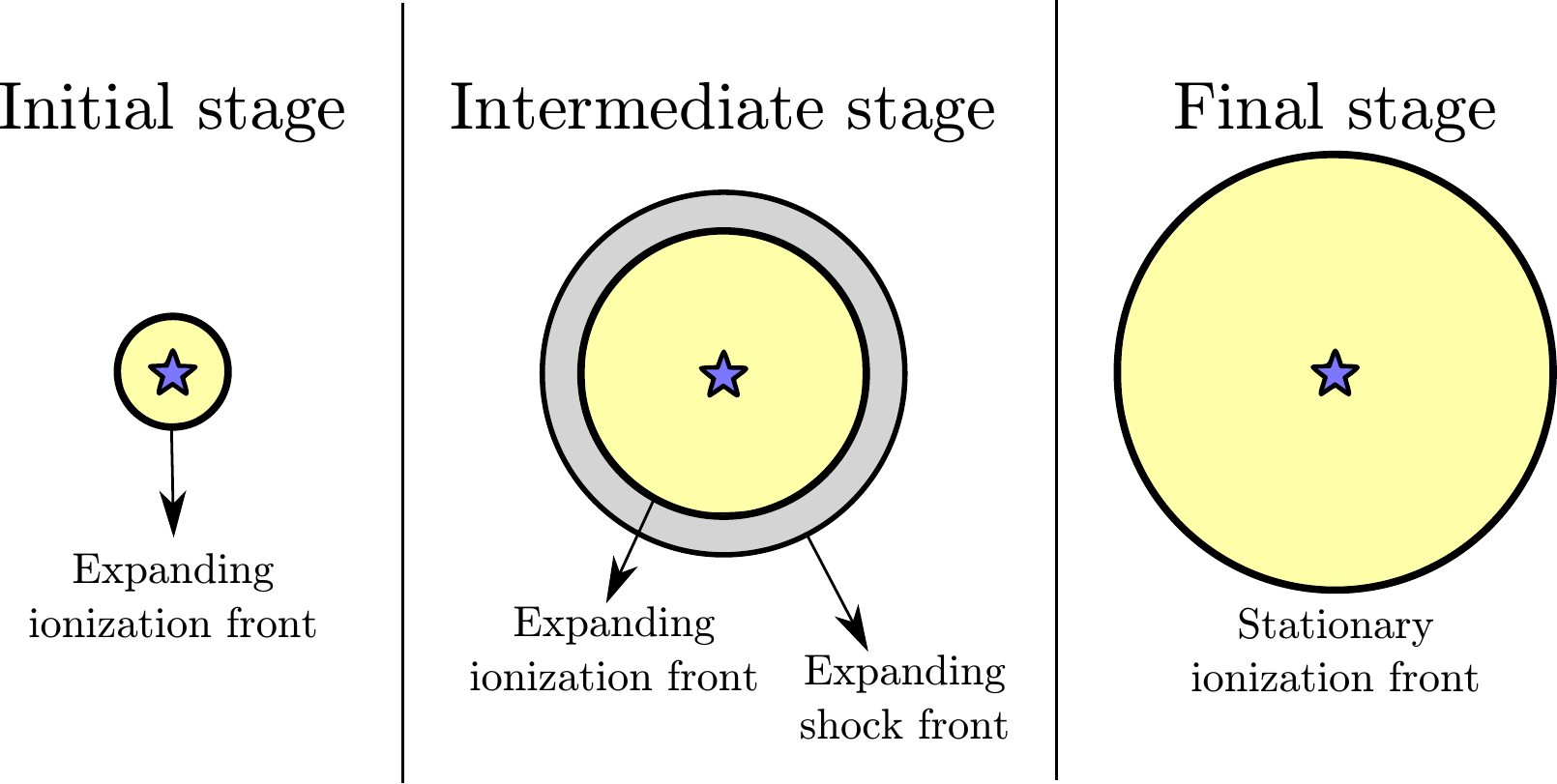}
  \caption{Three stages of \hii{} region evolution.  \textit{Left:}
    After a massive star forms, its UV radiation ionizes and heats the
    nearby interstellar medium, creating an \hii{} region.
    \textit{Middle:} After some time, the heated region expands and
    drives an outgoing shock (middle), carrying gas away from the
    star.  \textit{Right:} The \hii{} region comes into pressure
    equilibrium with the ambient medium.}
  \label{fig:hiiregion}
\end{figure}

A radiation source can only provide a finite number of ionizing
photons per second, so there must be some limit of ionizations
possible in the surrounding region.  Recombinations are happening
concurrently with ionizations in the highly ionized \hii~region.  In
some radial direction, if all of the ionizing photons are absorbed by
newly recombined atoms inside the \hii~region, there will be no more
flux left at the ionization front.  When this happens, the front
stalls out and reaches an equilibrium \citep{Stroemgren39}.  We can
determine the radius of the \hii~region $R_{\rm s}$, known as the
Str{\"o}mgren radius, by balancing the total number of ionizations and
recombinations in a region.  Assuming spherical symmetry and a static
and uniform medium, we set these total numbers to be equal and solve
for the radius.  It is larger for more luminous sources (higher
ionization rates) and for ambient gas with smaller densities (lower
recombination rates).

Figure \ref{fig:hiiregion} depicts the evolution of an \hii{} region
from when the star first ignites to the final stage when the
ionization front stalls out.  Initially the newborn star is surrounded
by a dense medium from which it formed.  The radiation front travels
near the speed of light at early times through this dense gas, where
the recombination rate is high.  The \hii{} region is now heated to
over 10,000~K and is encompassed by a cold ambient medium.  The gas
thus has a higher pressure than its surroundings and will start to
expand.  This marks the transition from the initial stage to the
intermediate stage after the time it takes a sound wave to cross the
\hii{} region.  As the material is forced away from the star by the
high pressure gas, it shocks with the ambient medium.  The shock wave
``sweeps up'' most of the gas in its path and accumulates mass,
leaving behind a more diffuse gas within the \hii{} region.  As the
density decreases, the recombination rate decreases accordingly.
Thus, the Str{\"o}mgren radius increases with time.  Eventually the
the \hii{} region comes into pressure equilibrium with the ambient
medium, and both the shock front and ionization front stall out at the
final Str{\"o}mgren radius.  This final equilibrium usually takes
hundreds of millions of years to manifest, which is much longer than
the lifetimes of massive stars, suggesting that this final stage is
usually not realized in nature.

More relevant for reionization, this scenario of a central source
ionizing its surrounding neutral gas can be extended to whole
galaxies.  Any ionizing radiation that escapes from the galaxy will
create a cosmological \hii{} region that is the building block of
reionization.

\section{Ending the Cosmic Dark Ages}
\label{sec:ending}

Cosmic reionization involves the coupling of non-linear physics of
galaxy formation with the non-local physics of gravity and radiation
transport to produce a global phase transition.  Reionization is
completely different from the local nature of the earlier phase
transitions that only depend on the thermal state of the plasma.  The
mixture of small- and large-scale physics makes for a complex problem.
Some important questions to ask about cosmic reionization are:
\textit{What are the main sources of reionization?  When does
  reionization begin and end?  What is the topology of the ionized
  regions?  What can be learned about early galaxies from reionization
  and vice-versa?  How does reionization affect galaxy formation?}

The present-day IGM has a mean temperature around $10^5 \unit{K}$ and
the primordial elements of hydrogen and helium are near complete
ionization \citep[e.g.][]{Dave01}.  Maintaining this relatively high
temperature and ionization state is an ultraviolet and X-ray radiation
background, sourced by countless galaxies and their central black
holes \citep{Haardt12}.  But how did the IGM become ionized and heated
in the first place?

\begin{figure}
  \centering
  \includegraphics[width=\textwidth]{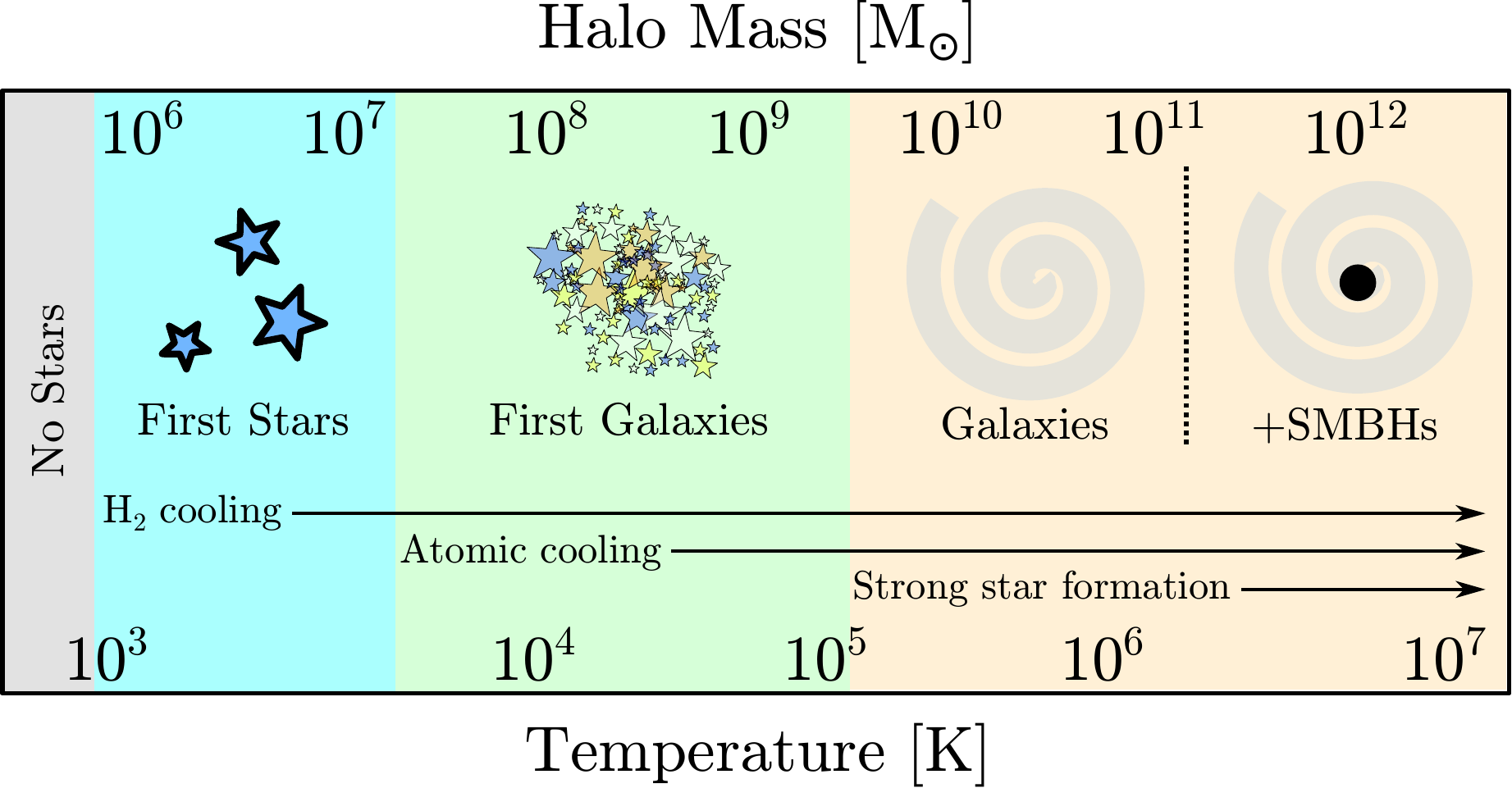}
  \caption{Below $10^{5.5}~\Ms$, halos cannot host cold gas and stars.
    As halos grow, they can host increasingly more cold gas and fuel
    stronger star formation, ranging from metal-free massive stars
    ($10^{5.5} - 10^7~\Ms$) to the first generation of galaxies ($10^7
    - 10^{9.5}~\Ms$) to more massive galaxies and supermassive black
    holes ($> 10^{9.5}~\Ms$).}
  \label{fig:halo_masses}
\end{figure}

This becomes a particularly rich question when combined with the fact
that galaxies were first assembling during the epoch of reionization
(EoR).  Depending on their mass, DM halos can support the formation of
various types of galaxies during the epoch of reionization, shown in
Figure \ref{fig:halo_masses}.  As halos grow with time, their
gravitational potentials deepen, and their gaseous components shock to
higher virial temperatures $T_{\rm vir}$.  The shocked gas undergoes
various radiative processes, cooling the gas.  Cold, dense gas fuels
star formation and is a good tracer of its strength.  Several physical
processes control the amount of cold dense gas, but two key processes
are the (1) efficiency of radiative cooling and (2) the ability of the
halo to retain gas that is heated by radiation and supernovae.  They
are both directly related to the halo mass.  Thus, we can categorize
halos by their mass and associate different types of behavior with
them.
\begin{itemize}
\item Very weak cooling and no star formation (Section \ref{sec:dark})
  below $\sim$$3 \times 10^5~\Ms$;
\item The first generations of massive primordial stars (Section
  \ref{sec:first-stars}) that are sparked by \hh{} cooling in halos
  with masses between $3 \times 10^5~\Ms$ and $10^7~\Ms$;
\item The first generations of galaxies (Section \ref{sec:first-gals})
  that form stars efficiently but its gas is susceptible to being
  blown out of the shallow gravitational potential in halos of masses
  between $10^7~\Ms$ and $3 \times 10^9~\Ms$;
\item More ``normal'' galaxies that exist in more massive halos.  They
  are reminiscent of a subset of present-day galaxies with strong star
  formation, is resistant to major gas losses, and may contain a
  central BH.
\end{itemize}

These star forming halos all contribute to cosmic reionization, which
can be divided into three phases \citep{Gnedin00}---pre-overlap,
overlap, and post-overlap---that describe the connectivity of the
cosmological \hii~regions, illustrated in Figure \ref{fig:timeline}.
In the pre-overlap phase, each galaxy produces its own \hii~region as
it forms.  These regions are divided by a vast neutral IGM, and their
evolution can be treated independently, only requiring the escaping
ionizing luminosity of the central galaxy.  In the overlap phase,
these regions start to combine with nearby regions \citep{Meiksin93}.
Multiple galaxies can contribute to the UV emissivity that pervades
the ionized region, dramatically increasing the mean free path of
ionizing photons and accelerating the reionization process.  Finally
in the post-overlap phase, most of the IGM is ionized with some
neutral patches remaining the universe.  These neutral regions erode
away as ionization fronts created from the UVB propagate into these
vestiges of an earlier cosmic time.

\subsection{The Dark Ages}
\label{sec:dark}

The Cosmic Dark Ages began after recombination at $z \sim 1090$
(380,000 years after the Big Bang) and lasted until the first stars
and galaxies started to light up the universe hundreds of millions of
years later.  However first, it is worthwhile to discuss the
supersonic relative velocities between DM and gas that arises before
and during recombination.  The importance of this phenomenon was
overlooked until recently \citep{Tselia10}.  Before recombination,
free electrons scatter off photons, strongly coupling the gas and
radiation, but the collisionless DM is not affected by this coupling.
These two components of the universe thus had different velocities as
radiation and gas decoupled at recombination.  The root-mean-square
value at this time was $\sim 30 \kms$ and fluctuated on comoving
scales between $\sim 3-200$~Mpc.  Thus on comoving scales of
$\sim$Mpc, the gas has a uniform bulk velocity relative to DM.  This
so-called streaming velocity decayed as $a^{-1}$, remained supersonic
throughout the Dark Ages, and prevented gas from collecting in the
potential wells of the smallest DM halos that would have otherwise
formed stars.

The mean gas temperature of the universe after recombination is
tightly coupled to the CMB temperature through Compton scattering
until a redshift $z_{\rm t} \approx 136$ at which time $T = 2.73
\unit{K} (1+z_{\rm t}) = 374 \unit{K}$ \citep{Peebles93}, where 2.73~K
is the current CMB temperature $T_{\rm CMB,0}$.  Afterwards in the
absence of any heating, the IGM in an expanding universe cools
adiabatically with its temperature decreasing proportionality to
$a^{-2}$, or equivalently $(1+z)^2$.  Throughout this epoch, DM halos
are continually assembling, but the gas cannot collapse into these
halos because of this excess thermal energy and kinetic energy from
streaming velocities. The Jeans mass
\begin{equation}
  \label{eqn:jeans}
  M_{\rm J} = \frac{\pi^{5/2}}{6} \frac{c_{\rm s}^3}{(G^3 \rho)^{1/2}}
\end{equation}
describes the required mass of an object that has enough gravity to
overcome thermal pressure, inducing a collapse.  Here $\rho$ and
$c_{\rm s} = (\gamma k_{\rm B} T / \mu m_{\rm p})^{1/2}$ are the gas
density and sound speed, respectively.  Using the IGM temperature
$T_{\rm gas}$ and the mean gas density $\bar{\rho}_{\rm b} = \rho_c
\Omega_b (1+z)^3$, one can calculate the cosmological Jeans mass
$M_{\rm J}$ that is around $32{,}000~\Ms$ at $z=30$ (100 Myr after the
Big Bang) and changes as $a^{-3/2}$.  It provides an estimate of the
minimum halo mass that can collect baryonic overdensities
\citep{Barkana01} in the case without streaming velocities.

\subsection{The first stars}
\label{sec:first-stars}

Historically, astronomers have categorized stars by their
metallicity\footnote{In astronomy, any element heavier than hydrogen
  (usually denoted by the variable X) and helium (Y) is historically
  termed as a metal (Z).} -- Population I for stars like our Sun,
which has 1.3\% metals by mass \citep{Asplund09}, and Population II
for stars with metallicities less than 1/10$^{\rm th}$ of the solar
metallicity fraction.  However, the Big Bang only produced hydrogen,
helium and trace amounts of lithium.  So there must have been an
initial population of stars composed of these light elements, whose
supernovae enrich later generations of stars with metals that we
observe today.

This first generation of stars, known as Population III (Pop III), is
inherently different than present-day stars because they form in a
neutral, pristine, untouched environment from a primordial mix of
hydrogen and helium.  Being metal-free reduces the cooling ability of
the collapsing birth cloud, resulting in stars that are typically more
massive than nearby stars \citep[e.g.][]{Bromm01, ABN02}.

Metal-free gas loses most of its thermal energy through \hh{}
formation in the gas-phase, using free electrons as a catalyst, in the
following reactions.
\begin{eqnarray}
  \textrm{H} + \textrm{e}^- &\rightarrow& \textrm{H}^- + \gamma\\
  \textrm{H}^- + \textrm{H} &\rightarrow& \textrm{H}_2 + \textrm{e}^-
\end{eqnarray}
Recombination leaves behind a residual free electron fraction on the
order of $10^{-5}$.  As gas falls into the halos, it shock-heats to
around the virial temperature $T_{\rm vir}$, and its electron fraction
is slightly amplified.

But \hh~is a fragile molecule that can be dissociated in the
Lyman-Werner (LW) bands between 11.1 and 13.6~eV at soft UV
wavelengths where the universe is optically thin.  Furthermore, the
intermediary product H$^-$ can be destroyed through the
photo-detachment of the extra electron that has an ionization
potential of 0.76~eV in the infrared.  Accordingly, the timing and
host halo masses of Pop III star formation is dependent on the
preceding star formation that produces the soft UV and infrared
radiation backgrounds.  The minimum halo mass that can support
sufficient \hh~formation that can induce a catastrophic collapse is
around $10^5~\Ms$ in the absence of LW radiation but steadily
increases to $10^7~\Ms$ in strong LW radiation fields
\citep{Machacek01}.  Additionally, streaming velocities can suppress
Pop III star formation in halos with masses $M \lsim 10^6 \Ms$ with
its exact value depending on the local streaming velocity magnitude
that varies on scales of tens of Mpc \citep[e.g.][]{OLeary12,
  Schauer19}.

Simulations of the first stars and galaxies that consider a LW
background have found that Pop III stars form at a nearly constant
rate of $\sim 3 \times 10^{-5}$~\sfr{} until the end of reionization
\citep[e.g.][]{Wise12}.  Each star produces a tremendous amount of
ionizing radiation because they are thought to be massive with
characteristic masses of tens of solar masses
\citep[e.g.][]{Hirano15}, some forming in binary systems and small
clusters \citep[e.g.][]{Turk09, Greif12}.  They have effective surface
temperatures $\sim 10^5 \unit{K}$ that is approximately
mass-independent above 20~\Ms{} because of the lack of typical
opacities associated with metals in their photospheres.  They live for
3--10~Myr and produce between $2 \times 10^{48}$ and $10^{50}$
hydrogen ionizing photons per second in the mass range 15--100 \Ms
\citep{Schaerer02}.  Most of these photons escape into the nearby IGM,
and the averaged escape fraction $f_{\rm esc}$ increases from 20\% for
a single 15~\Ms{} star to nearly 90\% for a single 200~\Ms{} star
\citep{Alvarez06}.  The exact values of $f_{\rm esc}$ will depend on
the total Pop III stellar mass in the halo, the halo mass, and its gas
fraction.  After their main sequence, some explode in supernova,
chemically enriching the surrounding few proper kpc, where the exact
fraction depends on their uncertain their initial mass function.
Because of this strong feedback, they quench their own formation,
blowing out most of the gas that originally existed in their host
halos.  Furthermore once the medium is enriched with metals, it's
'game over' for Pop III stars in affected regions because by
definition, they are metal-free.

\subsection{The first galaxies}
\label{sec:first-gals}

These heavy elements set the stage for the first galaxies that form in
larger halos, in which atomic (\lya) line cooling is efficient.
Accordingly, these halos are known as atomic cooling halos and have
virial temperatures above $\sim$8000~K.  These pre-galactic halos are
generally gas-poor ($f_{\rm gas} \equiv M_{\rm gas}/M_{\rm vir} \simeq
0.05-0.10$) because they are recovering from the gas blowout that
their halo progenitors experienced.  If the halo is below the atomic
cooling limit, star formation is bursty but still intense during
active periods, forming between $10^4$ and $10^5~\Ms$ of stars before
they can cool efficiently through atomic hydrogen transitions.  They
have star formation rates $\dot{M}_\star$ between $10^{-4}$ and
$10^{-3}$ \hsfr{}, doubling their stellar masses $M_\star$ every $\sim
30 \unit{Myr}$ (corresponding to a specific star formation rate,
$\mathrm{sSFR} \equiv \dot{M}_\star / M_\star \sim 3 \times 10^{-8}
\unit{yr}^{-1}$), and producing $\sim 3 \times 10^{49}$ ionizing
photons per second.  After the halo crosses the atomic cooling limit,
it can form stars in a continuous fashion at $\mathrm{sSFR} \sim 3
\times 10^{-8} \unit{yr}^{-1}$, which can vary by an order of
magnitude from galaxy to galaxy, depending on how it has been affected
by feedback from previous star formation \citep[e.g.][]{Kimm14,
  Kimm16, Xu16}.  By the time the halo mass reaches $10^9~\Ms$, the
first generations of galaxies contain between $10^6$ and $10^7~\Ms$ of
metal-poor ($Z \lsim 0.1~\zsun$) stars.

\begin{figure}[t]
  \centering
  \includegraphics[width=0.75\textwidth]{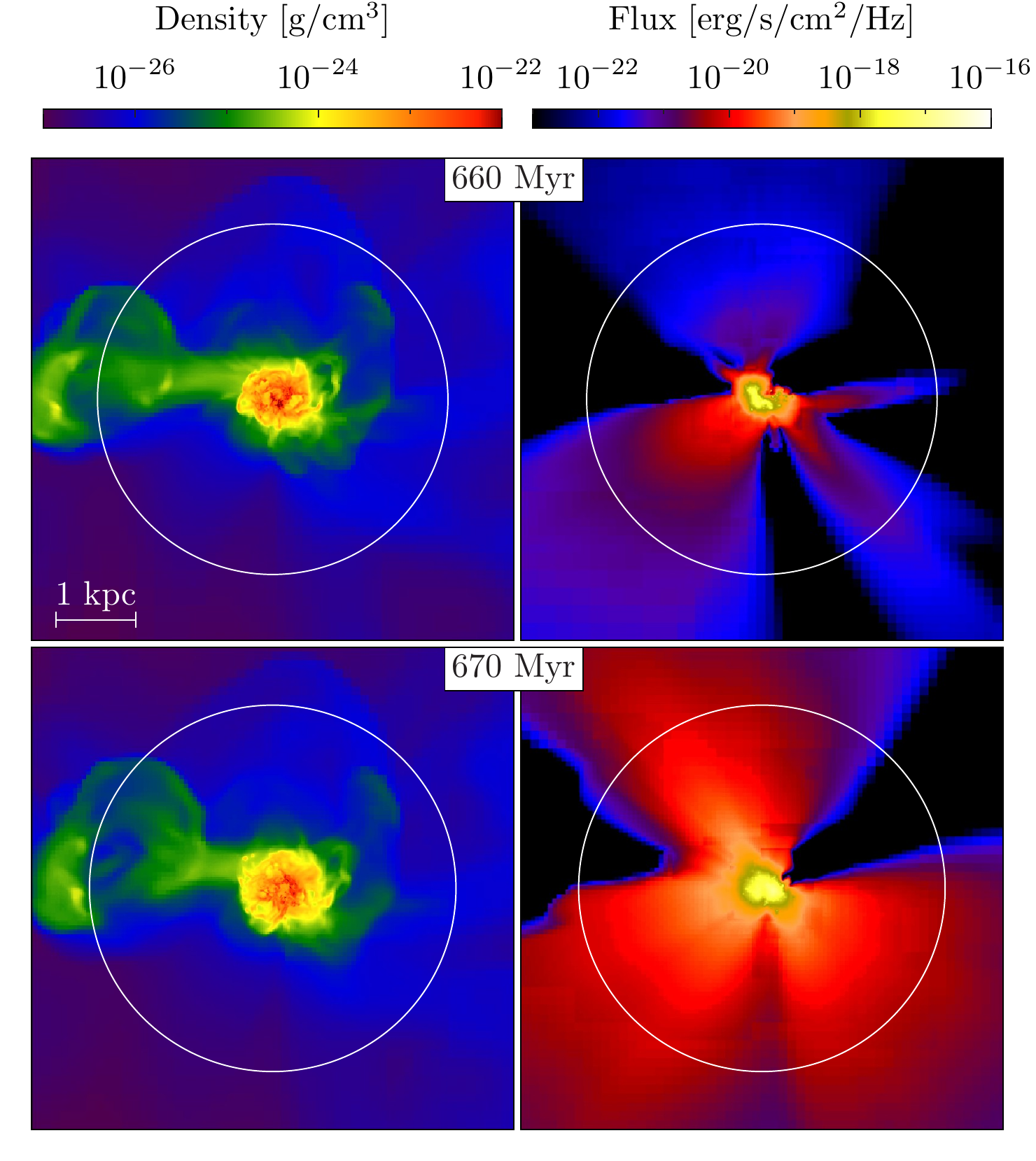}
  \caption{Projections of gas density (left) and UV radiation flux
    (right) of a first generation galaxy with a stellar mass $\sim
    10^5 \Ms$ at redshift $z \simeq 8$ with the bottom panels pictured
    10~Myr after the top panels.  The white circle marks the virial
    radius.  The UV escape fractions are 1\% (top) and 7\%
    (bottom). Adapted from \citep{Wise14}.}
  \label{fig:fesc}
\end{figure}

An important quantity in reionization calculations is the UV
escape fraction $f_{\rm esc}$, which is notoriously difficult to
observationally measure and to theoretically calculate.  Most
reionization models find that $f_{\rm esc} = 0.05-0.2$, independent of
halo mass, generally produce reasonable reionization histories
\citep[e.g.][]{Robertson13}.  In the past decade, there have been
great strides in the development of radiation hydrodynamics
simulations of the first galaxies in which a direct calculation of
$f_{\rm esc}$ is feasible.  This fraction is highly variable from
galaxy to galaxy, and even in a single object, it can vary from nearly
zero to unity over its formation sequence (see Figure \ref{fig:fesc}).
Because the interstellar medium (ISM) is clumpy, the ionization fronts
propagate outwards toward to the IGM at varying velocities with
respect to angle.  The ionizing radiation generally escapes in the
directions with small neutral column densities.  Once an ionized
channel is opened between a star cluster and the IGM, it remains
ionized as long as massive stars remain alive.  Thus, the value of
$f_{\rm esc}$ can be thought as the solid angular fraction that the
ionized channels cover.  Such efforts have found that the smallest
galaxies have high escape fractions.  The median time-averaged value
of $f_{\rm esc}$ is $\sim 0.5$ in halos with masses $M_{\rm vir}
\simeq 10^7~\Ms$, and it decreases to 0.05--0.10 at $M_{\rm vir}
\simeq 10^8~\Ms$ \citep{Kimm14, Ma15, Xu16}.  When the total escaping
photons are integrated over all galaxies, half of the photon budget to
reionization originate from halos with $M_{\rm vir} \lsim 10^9~\Ms$.

\subsection{The first black holes}

Black holes grow through mergers of two black holes and the accretion
of gas.  The latter is the primary growth mechanism, where material
falls down the deep gravitational potential well.  It forms an
accretion disk orbiting around the black hole because of conservation
of angular momentum.  Eventually, gas migrates to increasingly smaller
radii and falls into the black hole.  In the process, the gas is
heated intensely from the strong gravity field and emits radiation.
Any outward radiation interacts with the inflowing gas, which can
suppress the growth rate of the black hole.

Because Pop III stars are thought to be massive, a large fraction will
leave a black hole remnant.  Temperatures in accretion disks around
stellar-mass black holes are between $10^4$ and $10^7 \unit{K}$ and
thus emits strongly in the hard UV and X-ray energies.  Their UV
luminosities are insignificant when compared to stellar sources, but
their X-rays should have an impact on the thermal and ionization state
of the IGM.  These high energy photons can penetrate much deeper into
the IGM, creating large partially ionized regions with $x_e =
0.01-0.02$ out to distances of 100~kpc \citep{Xu14}, creating a much
different reionization topology than stellar sources
\citep{Mesinger13}.

Pop III stars leave behind some of the first black holes in the
universe.  These could be the ``seeds'' for more massive black holes
that we observe in the nearby Universe.  They can possibly grow to
supermassive black holes that are millions and sometimes billions of
times more massive than our Sun.  The most extreme cases have masses
over $10^9~\Ms$ at redshifts $z > 6$ when the universe was younger
than 800 million years old \citep{Banados18}.  The formation and growth of
black holes are still active research topics.  A few pressing
questions are: \textit{How did massive black holes in the distant
  universe grow so rapidly?  Did their radiation contribute to
  reionization?  How often did stellar-mass black holes grow into
  supermassive black holes?  Where are these stellar remnants today?}
Observations from the epoch of reionization will provide us with clues
in order to solve these mysteries.

\section{Constraints from the Edge of the Observable Universe}

Now that we have covered the basic astrophysics and the theoretical
aspects of reionization, we can gain insight on cosmic reionization
through some key observations.  There are various independent probes
of reionization, either measuring the ionization and thermal state of
the IGM or the properties of galaxies forming during the EoR.  The
vast majority of these observations are difficult with the photons
streaming across the observable universe to Earth, requiring long
exposure times and/or large telescopes.  Here we cover the basic
physics behind each method and the latest constraints.

\subsection{QSO Spectra}

\begin{figure}
  \centering
  \includegraphics[width=\textwidth]{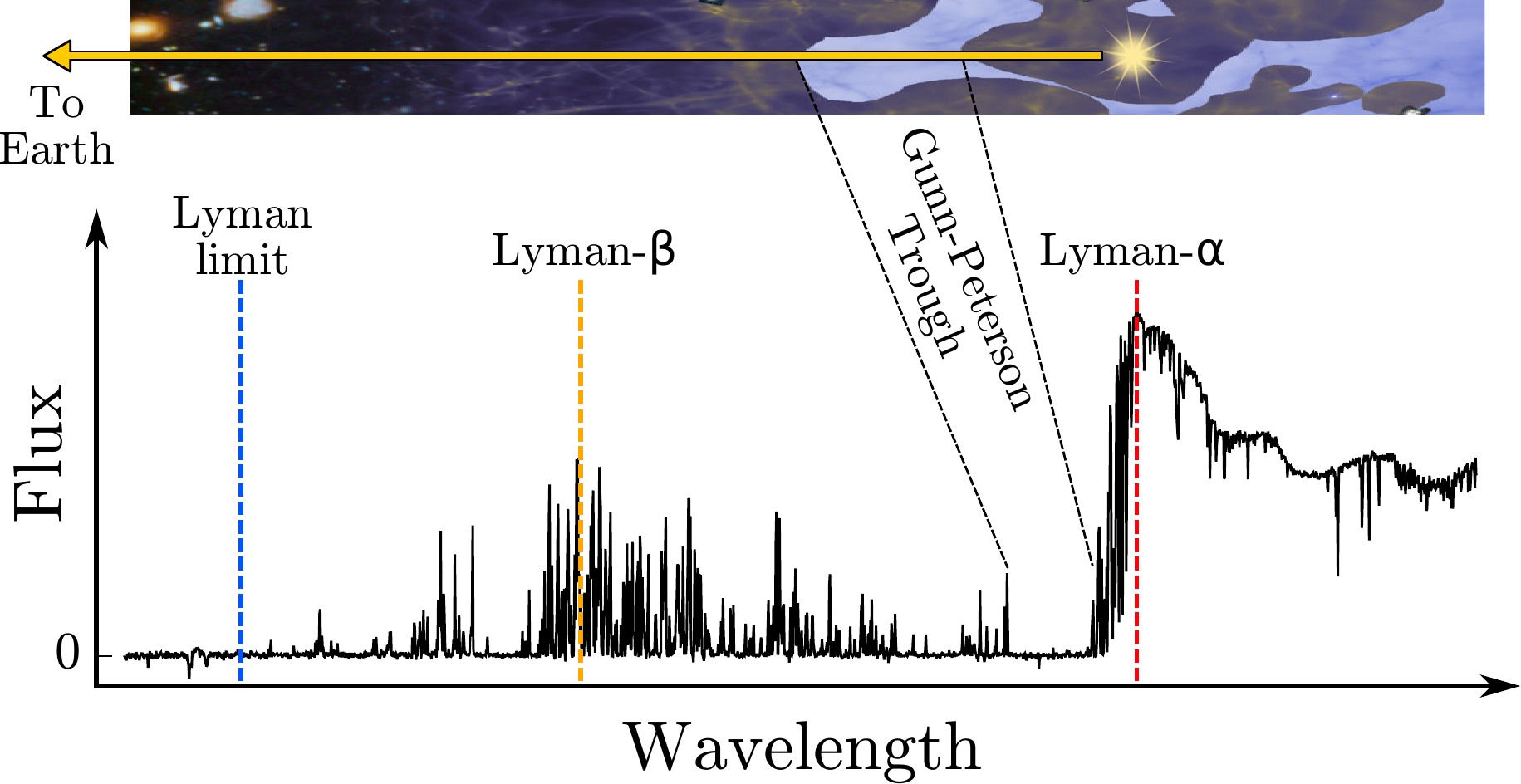}
  \caption{Light from distant quasars, powered by supermassive black
    holes, can probe the ionization and thermal state intergalactic
    medium.  Overdense clumps of intergalactic gas absorb some
    fraction of light from the intrinsic spectrum (bottom) when the
    photons ionize any neutral hydrogen.  Only lines in the Lyman
    series, down to the Lyman limit (912~\AA), with Ly$\alpha$
    (1215~\AA) and Ly$\beta$ (1026~\AA) being the strongest.
    Absorption lines from clouds at various redshifts create a
    Ly$\alpha$ forest.  When these lines becomes numerous enough, it
    creates a Gunn-Peterson trough that is indicative of an elevated
    neutral hydrogen fraction and the end of the Epoch of
    Reionization.}
  \label{fig:qal}
\end{figure}

Light from distant QSOs can be used as flashlights to probe the
intervening IGM.  Neutral gas clouds at some redshift $z_{\rm abs}$
will strongly absorb light at the Lyman-$\alpha$ wavelength
$\lambda_\alpha = 1216 \unit{\AA}$ in its rest frame that is then
redshifted to a wavelength $\lambda_\alpha (1+z_{\rm abs})$ in
observations.  The schematic and accompanying spectrum in Figure
\ref{fig:qal} depicts this technique of quasar absorption spectra
mapping out the IGM.  It shows that many gas clouds at varying
redshifts between the QSO and us has blocked out much of the
radiation at wavelengths shorter than $\lambda_\alpha$.

As discussed previously, Gunn and Peterson used QSO spectra to
determine that the IGM must be highly ionized, otherwise an absorption
trough, now termed the Gunn-Peterson (GP) trough, would have been
present at $\lambda_\alpha < 1216 \unit{\AA}$.  The optical depth to
\lya{} photons ($n = 1 \rightarrow 2$) is extremely high, upwards of
$10^5$ in a completely neutral IGM at high redshifts.  Because optical
depth is proportional to neutral hydrogen density, the
IGM even with a tiny neutral fraction $f_{\rm HI} \equiv n_{\rm
  HI}/n_{\rm H} \sim 10^{-4}$ will be optically thick to \lya{}
photons and absorb all light at a wavelength $1216 (1+z_{\rm
  abs})$~\AA.  Because \lya{} is such an effective absorber, this
method only probes the neutral fraction of highly ionized gas and
cannot be used to peer into the EoR.

There have been dozens of $z \ge 6$ QSO absorption spectra that have
GP troughs \citep{Fan06b}, which are increasingly opaque with
redshift.  This is a strong constraint that cosmic reionization is
complete by $z \sim 6$.  The IGM near QSOs are also exposed to a
strong radiation field and should be more ionized than the typical
IGM.  This is known as the proximity effect, resulting in weaker
absorption between the \lya{} emission line and GP trough.  The sizes
of these ionized regions are on the order of tens of Mpc at $z \sim 6$
\citep[e.g.][]{Alvarez09}.  Similar to this type of analysis,
absorption from a partially ionized IGM in the proximity zone will
produce a damping \lya{} wing that occurs when the neutral column
density $N_{\rm HI} \gsim 10^{20} \unit{cm}^{-2}$.  The spectrum of
one of the most distant QSOs at $z=7.08$ constrains the ionized
fraction to be $x_{\rm e} \simeq 0.60$ at this redshift
\citep{Greig16_QSOz7}.  On the opposite end, there are still some
neutral islands with sizes up to 160~Mpc at $z = 5.5-6$
\citep{Becker15}, but they quickly erode as the increasing UV
background ionizes them \citep{Fan06b}, showing that cosmic
reionization is an inhomogeneous process.

\subsection{The \lya{} forest}

The \lya{} forest denotes the myriad of \lya{} narrow absorption lines
coming from clouds in the IGM between the quasar and us.  For a cloud
existing at a redshift $z$, they will create an absorption line at
wavelength $1215(1+z)$\AA.  They become more abundant with increasing
redshift \citep{Songaila04} and probe clouds with column densities
$\log(N_{\rm HI}/\textrm{cm}^{-2}) = 12-16$.  These lines become so
abundant that they start to block out all of the background light,
transforming into a GP trough at $z \sim 6$.  The example spectrum in
Figure \ref{fig:qal} shows a dense \lya{} forest, transmitting very
little light at wavelengths between \lya{} and Ly$\beta$ (1026~\AA).
This particular spectrum transmits more light at shorter wavelengths,
or equivalently, lower redshifts, suggesting that this line of sight
is becoming more ionized with decreasing redshift.  One constraint on
the ionized fraction is the 'dark fraction' of QSO spectra in the
\lya{} forest that originate from either neutral patches or residual
neutral hydrogen in ionized regions \citep{Mesinger10}.  Taken at $z =
5.9$, the dark fraction in the \lya{} forest results in an lower limit
of $x_{\rm e} > 0.94$ \citep{Greig16}.

Because \lya{} forest clouds contain such small column densities, they
are prone to ionization and heating from the ultraviolet background
(UVB) produced by galaxies and quasars, and thus are excellent
thermometers of the post-reionization universe.  The UVB can be
quantified by the hydrogen ionization rate
\begin{equation}
  \Gamma(z) = 4\pi \int_{\nu_{912}}^\infty J_\nu(z) \sigma_{\rm
    HI}(\nu) \frac{d\nu}{h\nu},
\end{equation}
that comes from the sum of ionizing photons that interact with a
neutral hydrogen atom.  Here $\sigma_{\rm HI}$ is the photoionization
cross-section, $J_\nu$ is the specific intensity, and $\nu_{912} =
3.28 \times 10^{15} \unit{Hz}$ is the frequency of the Lyman limit.
The ionization rate can be derived from the \lya{} forest lines,
provided its temperature and optical depth to \lya{} photons.  The
temperature can be calculated from \lya{} forest line widths, which
are affected by the Doppler effect of thermal motions of neutral
hydrogen in the clouds and the thermal smoothing of the absorber over
the time it has been exposed to the UVB.  The optical depth can be
calculated in photoionization equilibrium \citep{McDonald01}.  Data
from the Sloan Digital Sky Survey \citep[SDSS;][]{York00} has shown
that the ionization rate is relatively constant between $z=2-5$ and
sharply increasing with time between $z=5-6$ \citep{Becker13}.  The
ratio of ionizing to non-ionizing radiation increases by a factor of
$\sim 3$ going from $z=3$ to $z=5$, suggesting that galaxies are more
efficient producers of ionizing photons at earlier times.  Lastly, the
sharp evolution in $\Gamma$ at $z > 5$ could be caused by either an
increase in ionizing emissivity from galaxies and black holes or the
opacity of the IGM.  The latter decreases after EoR as dense neutral
clouds are photo-evaporated by the UVB, increasing the mean-free path
of ionizing photons.

The thermal history of the IGM, probed by the \lya{} forest also
places constraints on the ionizing source spectra.  After the cloud
has been heated by some radiation source, it actually never reaches thermal
equilibrium.  We can consider a thermal model with the UVB as a
heating source and adiabatic expansion as the coolant.
Several groups have found that the typical IGM temperature is $\sim
10^4 \unit{K}$ at $z=5$ \citep[e.g.][]{Schaye00, Becker13}.  The
low-density IGM has a long cooling timescale and, thus, has a thermal
memory of reionization.  The exact thermal evolution depends on the
timing of the initial photoheating, giving the time available to cool
to $\sim 10^4 \unit{K}$ at $z = 5$ and the spectral hardness of
ionizing sources \citep{Hui97, Bolton10}.  At later times, the double
ionization of helium increases the mean IGM temperature to $3 \times
10^4 \unit{K}$ at $z = 2$ \citep[e.g.][]{Bolton12}.

\subsection{Cosmic microwave background}

\begin{figure}
  \centering
  \includegraphics[width=\textwidth]{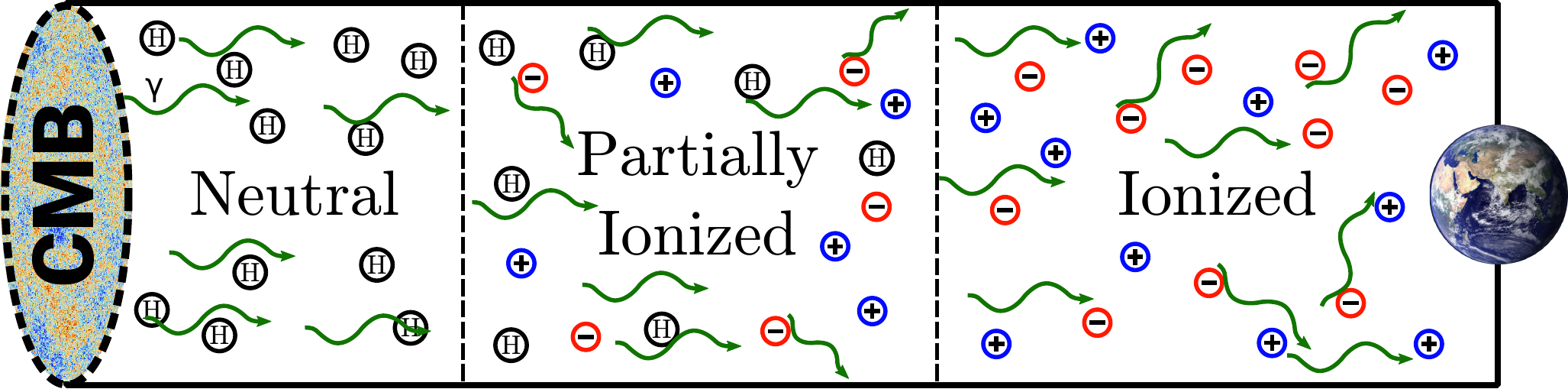}
  \caption{Photons from the CMB can be scattered by free electrons,
    called Thomson scattering, as they propagate to detectors on
    Earth.  Only the total amount of Thomson scattering can be
    measured from the CMB, which is related to the timing of cosmic
    reionization.}
  \label{fig:thomson}
\end{figure}

While QSO absorption spectra probe the end of cosmic reionization, the
CMB photons travel from the surface of last scattering to Earth and
may scatter off free electrons, which is depicted schematically in
Figure \ref{fig:thomson}.  Thomson scattering polarizes the CMB at
large angular scales, resulting in a Thomson scattering optical depth
$\tau_{\rm es}$
that is directly related to the column density of free electrons.
This measure is an integrated one and tells us little about the
reionization history and only about the approximate timing of
reionization.  
A fully ionized IGM between $z=0$ and $z=6$ results in $\tau_{\rm es}
= 0.039$, and the remaining portion ($z>6$) of the integral depends on
the reionization history $\bar{x}_{\rm e}(z)$.  The most recent Planck
2018 \citep{Planck18_Cosmo} measurement of $\tau_{\rm es} = 0.0561 \pm
0.0071$, corresponding to a reionization redshift $z_{\rm re} = 7.82
\pm 0.71$ when the universe was half ionized ($\bar{x}_{\rm e} =
0.5$).

\subsection{Neutral hydrogen (21 cm) absorption and emission}
\label{sec:21cm}

\begin{figure*}
  \centering

  \subfigure[Rendering of the final configuration of the Hydrogen
  Epoch of Reionization Array (HERA) with 331 antennas in South
  Africa.]
  {\raisebox{0.2\width}{\resizebox*{0.44\textwidth}{!}
      {\includegraphics[width=0.44\textwidth]{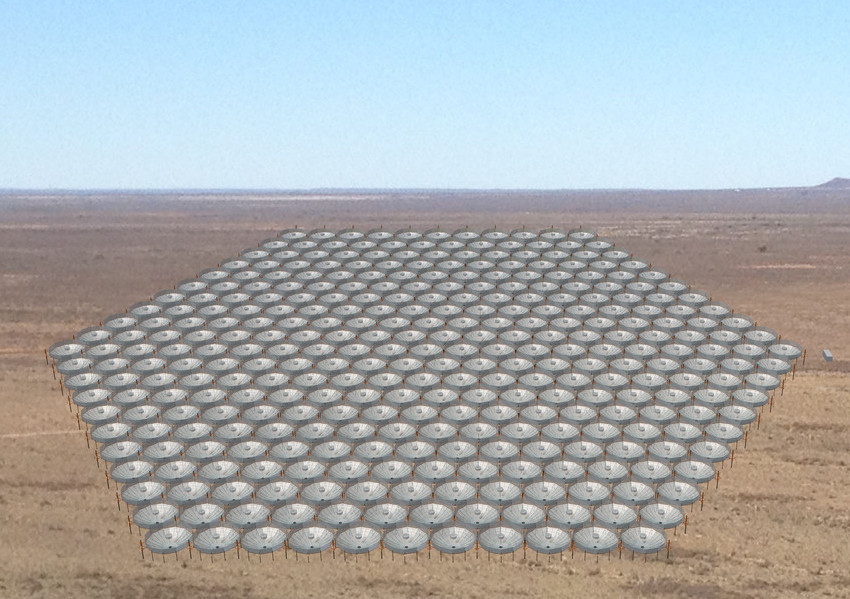}}}}
  \hfill
  \subfigure[Rendering of the Square Kilometer Array (SKA) to be
  co-located in Australia and South Africa.]
  { \resizebox*{0.5\textwidth}{!}
    {\includegraphics[width=0.5\textwidth]{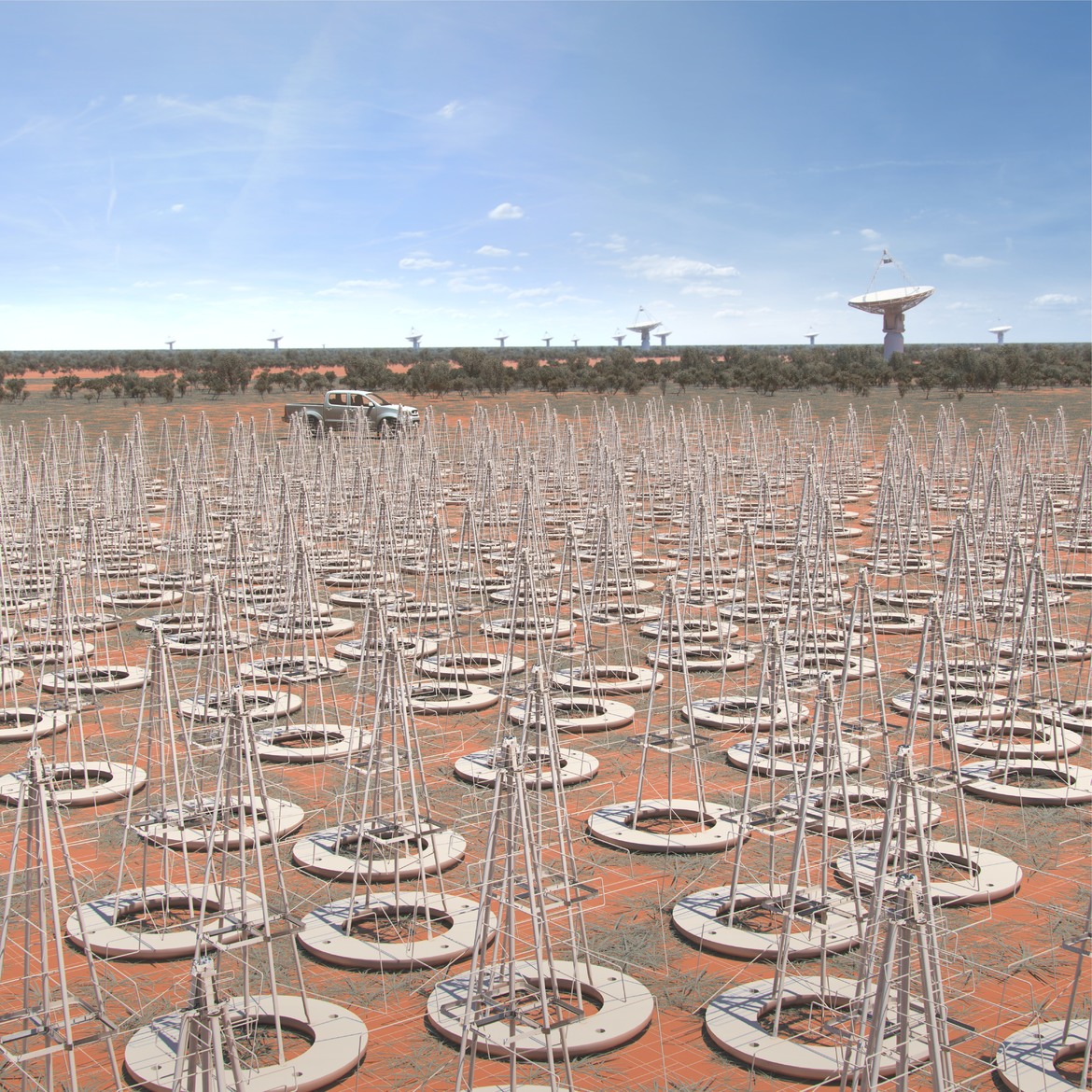}}}
  \caption{Next generation 21-cm radio telescopes that will directly
    explore the Epoch of Reionization.}
  \label{fig:21cm-obs}
\end{figure*}

Perhaps the most direct measure of cosmic reionization comes from
neutral hydrogen emission of the hyperfine splitting of the ground
state, the 21-cm transition ($\nu = 1420.4 \unit{MHz}$; $E_{21} = 5.87
\times 10^{-6} \unit{eV}$).  The hydrogen atom has a slightly lower
energy when the spins of its proton and electron are anti-parallel
than when they are parallel.  The first detection of extraterrestrial
21-cm emission from neutral hydrogen happened in 1951 \citep{Ewen51,
  Muller51}, and it was not until four decades later that it was
realized that 21-cm observations could be used to probe reionization
\citep{Scott90}.

When the IGM is mostly neutral, the universe is glowing in this
radiation, and most of it is not absorbed as it travels toward Earth.
Its detection is complicated by astrophysical foreground and
terrestrial sources, especially considering that the redshifted 21-cm
emission is at $101.5 [(1+z)/14]^{-1} \unit{MHz}$, squarely in the FM
band.  Before any stars form, the spin temperature
\begin{equation}
  T_{\rm s} = T_\ast \ln\left( \frac{3n_0}{n_1} \right)
\end{equation}
measures the relative occupancy of the electron spin levels.  Here
$T_\ast = E_{21}/k_{\rm B} = 0.0682 \unit{K}$, and $n_0$ and $n_1$ are
the singlet and triplet hyperfine levels of the ground state.  The
factor of 3 comes from the ratio of statistical weights between these
levels.  The spin temperature tightly couples to the CMB temperature
$T_{\rm CMB} = 2.73 (1+z) \unit{K}$ because the gas density is not
high enough to couple with the kinetic temperature $T_{\rm K}$ that
cools as $a^{-2}$ from adiabatic cosmic expansion.

Absorption or emission by neutral hydrogen changes the 21-cm
(differential) brightness temperature $\delta T_{\rm b}$, which is the
spin temperature relative to the background (CMB) temperature.
Because the radiation is emanating from the neutral component of the
IGM, it is usually multiplied by its neutral fraction $(1 -
\bar{x}_{\rm e})$.
Positive and negative values denote emission and absorption at 21-cm.
In addition to the neutral fraction of the IGM, X-ray and \lya{}
radiation can modify the 21-cm signal.  \lya{} radiation effects
become dominate after $z \sim 30$ as the first stars begin to form,
driving a {\it decrease} in $\delta T_{\rm b}$.  Then the IGM begins
to be partially ionized and heated by X-ray sources, {\it increasing}
$\delta T_{\rm b}$.  Eventually the IGM becomes ionized by UV sources,
causing the brightness temperature $\delta T_{\rm b}$ to asymptote to
zero as $\bar{x}_{\rm e}$ approaches unity.  In summary, the 21-cm
$\delta T_{\rm b}$ signal would appear a trough that deviates from
zero and should smoothly vary because it is a volume average over the
Universe.

A measurement of the brightness temperature evolution would place
strong constraints on the reionization history and the nature of the
ionizing sources.  In particular, the location of the trough in
$\delta T_{\rm b}$ relays information about the \lya{} and X-ray
emissivities of the first stars, black holes, and galaxies.  Bowman et
al. \citep{Bowman18_EDGES} reported on the first detection of such an
absorption trough with the Experiment to Detect the Global Epoch of
Reionization Signature (EDGES).  It is centered at 78~MHz,
corresponding to a redshift $z \sim 18$, and has anomalously sharp
edges and a strong amplitude.  First, its timing suggests that early
star formation must have been intense in low-mass ($M_{\rm h} \gsim
10^{10}~\Ms$) halos \citep{Mirocha19}.  Second, its shape indicates a
rapid coupling of the spin temperature to the gas temperature and was
not predicted by prior models \citep{Cohen17}.  Kaurov et
al. \citep{Kaurov18} argue that rare and massive galaxies at $z \sim
15-20$ are predominately responsible for this signal.  Last, the
absorption trough is consistent with a cold IGM prior to reionization.
Perhaps the most mysterious aspect of the EDGES detection is its
extreme depth, which suggests that the IGM is even colder than an
adiabatically cooling IGM.  This requires an additional cooling at
very high redshifts ($z = 30-100$) could indicate new physics, such as
interactions between baryons and dark matter \citep{Barkana18}.

There are several experiments also aiming for the same detection,
which include PAPER \citep{paper}, LOFAR \citep{lofar}, MWA
\citep{mwa}, HERA \citep{hera}, and SKA \citep{ska}.  It is essential
to independently confirm this groundbreaking direct detection of
reionization and to obtain additional observations of this cosmic
phase transition.

%\li See the reviews of Finlator (2012) and Natarajan+ (2014)

\subsection{Sources of reionization}

From the direct and indirect IGM observations just discussed, we know
that cosmic reionization occurred between $z = 6-15$.  But what
sources were responsible for producing the required ionizing radiation
for such a phase transition?  QSOs are some of the brightest objects
in the universe, but their number densities are not high enough
\citep[e.g.][]{Kashikawa15} to significantly contribute to the UVB and
the overall photon budget of reionization.  The latest studies have
shown that they only contribute 1--5\% of the photon budget at $z = 6$
\citep[e.g.][]{Willott10, Grissom14}, however see Madau \& Haardt
\citep{Madau15} for a counterpoint.  This leaves starlight from
galaxies to propel reionization.  Two important questions about the
characteristics of high-redshift galaxies are: \textit{How abundant
  are galaxies as a function of luminosity and redshift?  How many
  ionizing photons escaped from these galaxies into the IGM?}  The
first question is addressed by counting galaxies and computing a
luminosity function (LF), and the second is a harder quantity to
measure as a neutral IGM is opaque to ionizing photons and needs to be
inferred from their UV continuum redward of \lya{}.

Recent observational campaigns have provided valuable constraints on
the nature of the first galaxies, their central BHs, and their role
during reionization.  In the rest-frame UV, the Hubble Space Telescope
(HST) {\it Ultra Deep Field} \citep{Ellis13} and {\it Frontier Fields}
campaigns \citep{Coe15} can probe galaxies with stellar masses as
small as $10^7~\Ms$ at $z \gsim 6$ and as distant as $z \simeq 11$
\citep{Laporte16, Oesch16}.  The LF is best described with a Schechter
fit \citep{Schechter76} as a function of luminosity,
\begin{equation}
  \phi(L) = \phi^* \left( \frac{L}{L^*} \right)^\alpha \, \exp\left(
    -\frac{L}{L^*} \right)
\end{equation}
that gives how many galaxies exist per comoving Mpc$^3$ per decade of
luminosity. As a function of absolute magnitude $M$, it reads as
\begin{equation}
  \phi(M) = \frac{\ln(10)}{2.5} \phi^* 10^{0.4(M-M^*)(\alpha+1)}
  \exp\left[ -10^{-0.4(M-M^*)} \right].
\end{equation}
The LF exponentially decays at the bright-end and is a power law at the
faint-end.  There are three parameters in this fit: the characteristic
luminosity $L^*$ (or magnitude $M^*$) that denotes the transition
between power-law and exponential decay, the number density
normalization $\phi^*$, and faint-end slope $\alpha$.  For the
purposes of reionization sources, the faint-end is the most relevant
because these faint galaxies should be very numerous.  However the
luminosity function should flatten and decrease at very low
luminosities because galaxies cannot form in small halos (see
Fig. \ref{fig:halo_masses}).  The faint-end slope and the luminosity
at which the LF flattens is key when computing the total number
density of galaxies and their ionizing emissivity.  Various groups
have constrained $\alpha \simeq -2$ at $z \ge 6$
\citep[e.g.][]{McLure13}, and there are slight hints from the {\it
  Frontier Fields} that the LF flattens above a UV absolute magnitude
of --14 \citep{Livermore16, Bouwens16}.  Based on this steep slope,
there should be an unseen population of even fainter and more abundant
galaxies that will eventually be detected by next-generation
telescopes, such as JWST (James Webb Space Telescope) and 30-m class
ground-based telescopes.

The ionizing emissivity ($\rho_{\rm UV}$; in units of erg s$^{-1}$
Hz$^{-1}$ Mpc$^{-3}$) is a key quantity in reionization calculations.
Ionizing radiation is extremely difficult to observe, so we have to
estimate it from the non-ionizing portion of galaxy spectra.  Given a
relation between total luminosity $L$ and star formation rate (SFR),
we can integrate the product of SFR and the LF over luminosity to
obtain a SFR density (in units of $\Ms \unit{yr}^{-1}
\unit{Mpc}^{-3}$).  This quantity can then be converted into the
ionizing emissivity $\rho_{\rm UV}$ with two factors.  The first is
the number of ionizing photons emitted per stellar baryon $f_\gamma
\simeq 4000-13000$, which depends on stellar metallicity.  The second
and most uncertain is the fraction $f_{\rm esc}$ of ionizing photons
that escape into IGM.  Finkelstein et al. \citep{Finkelstein12} used
\lya{} forest observations to place an upper limit on the average
$\langle f_{\rm esc} \rangle < 0.13$ at $z = 6$.  However this does
not prevent the average $f_{\rm esc}$ from being larger at higher
redshifts \citep{Robertson13}.  Direct measurements of $f_{\rm esc}$
is impossible during EoR because the IGM optical depth in the \lya{}
forest only drops to unity at $z \sim 3$ as these absorption systems
become less abundant with time.  Nevertheless, deep narrow-band galaxy
spectroscopy and imaging have detected Lyman continuum emission in
numerous $z \sim 3$ galaxies with $f_{\rm esc}$ values ranging from an
upper limit of 7\%--9\% for bright galaxies \citep{Siana15} to
10\%--30\% for fainter \lya{} emitters \citep{Nestor13} and $33 \pm
7\%$ for 'Lyman-continuum galaxies' \citep{Cooke14}.  These
measurements combined with theoretical calculations may be taken as a
guide to the $f_{\rm esc}$ values expected from EoR galaxies.

\subsection{Summary of observations}

Having reviewed the primary observational constraints on cosmic
reionization, there are clearly several independent measures of this
grand event.  The end of reionization is probed by QSO absorption
spectra. The \lya{} forest and GP troughs are used to constrain the
residual neutral fraction ($\simeq 10^{-4}$ at $z=6$) and remaining
neutral patches.  The thermal state of the \lya{} forest retains some
'fossil memory' of its initial heating during reionization and can be
used to constraint the timing and heating source spectrum.
Furthermore, the evolution of the UVB is determined from the \lya{}
forest and is steeply increasing with time at $z > 5$.  The CMB
provides both an integral measure of reionization with the universe
being half-ionized at $z_{\rm re} = 7.82 \pm 0.71$.  Future 21-cm
experiments will be able to further constrain the reionization history
and the nature of the ionizing sources.  From the measurement of the
steep faint-end LF slope $\alpha \simeq -2$ at $z \ge 6$, young
star-forming galaxies are most likely the primary drivers of
reionization.

\section{Reionization Modeling}

\begin{figure}
  \centering
  \includegraphics[width=\textwidth]{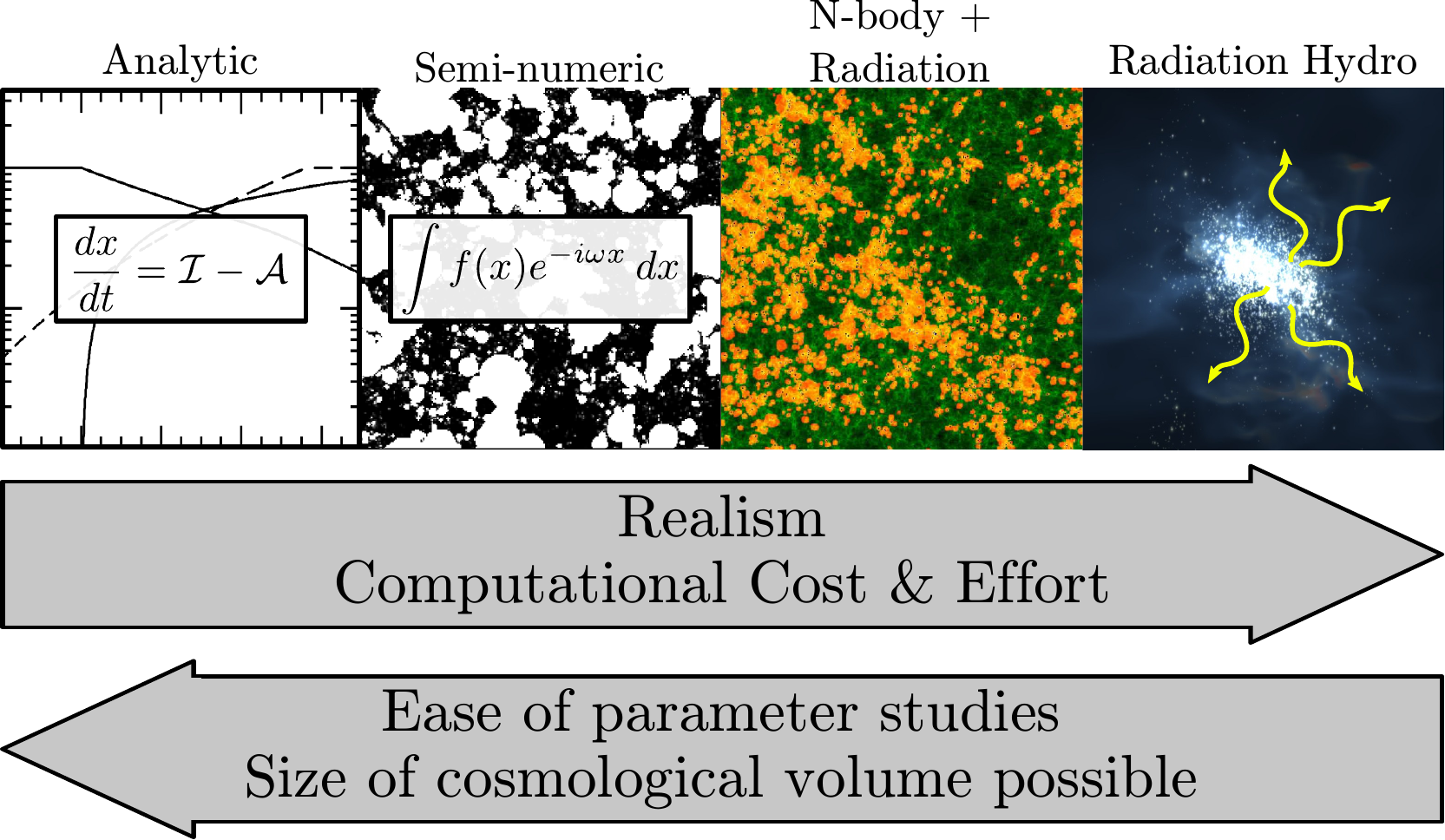}
  \caption{Four different numerical models that capture the process of
    cosmic reionization: volume-averaged analytic models;
    spatially-dependent semi-numeric models; radiative transfer
    calculations that use matter distributions from N-body
    calculations; full radiation hydrodynamic galaxy formation
    simulations.  The increased physicality and realism comes at the
    price of added computational cost, limiting the ability to explore
    different reionization models and the associated parameter space.}
  \label{fig:methods}
\end{figure}

Cosmic reionization can be modeled in a variety of methods, which are
depicted in Figure \ref{fig:methods}.  They range from a
volume-averaged approach, taking less than a second to compute, to
full radiation hydrodynamics cosmological simulations of galaxies and
the IGM, taking months to complete.  These methods have their
advantages and disadvantages, and they complement each other.  They
use the appropriate observational constraints to make predictions
about the reionization history, the nature of its sources, and future
observational techniques on further constraining this epoch.

\subsection{Volume-averaged analytical models}

On the most basic level, reionization models count the number of
ionizing photons and compare it with the number of hydrogen atoms in
some specified volume, which was first (to the best of the author's knowledge)
studied by Arons \& McCray \citep{Arons70} in a time-dependent
approach.  As we have covered in the basic physics of ionization
balance, recombinations can still occur in a strong ionizing radiation
field, and thus cosmic reionization requires more than one ionizing
photon per hydrogen atom.  The rate of change in the ionized fraction
$x_{\rm e}$ of the universe can be computed by considering the
comoving photon emissivity $\dot{n}_\gamma$ and recombinations in the
ionized regions,
\begin{equation}
  \frac{dx_{\rm e}}{dt} = \frac{\dot{n}_\gamma}{\bar{n}_{\rm H}} -
  \frac{x_{\rm e}}{\bar{t}_{\rm rec}},
\end{equation}
This differential equation is integrated from a neutral medium
($x_{\rm e} = 0$) at very early times before any UV sources form ($z
\sim 50-100$) until the universe is completely ionized ($x_{\rm e} =
1$).  Its evolution gives a reionization history, which then can be
integrated to obtain the Thomson scattering optical depth $\tau_{\rm
  es}$.

Here $\bar{n}_{\rm H}$ is the mean comoving hydrogen number density,
and $\bar{t}_{\rm rec} = [C \alpha_{\rm B} \bar{n}_{\rm H}(1 +
Y/4X)(1+z)^3]^{-1}$ is an effective recombination time.  Recall that
$X = 0.76$ and $Y = 1-X$ are the hydrogen and helium number fractions,
respectively.  The clumping factor $C \approx 3-5$ during EoR accounts
for enhanced recombinations in a clumpy ionized IGM
\citep[e.g.][]{Pawlik09, So14}.  There are various definitions for the
clumping factor \citep[see][for a discussion]{Finlator12}, but the
most straightforward definition is $C \equiv \langle \rho^2 \rangle /
\langle \rho \rangle^2$ and is restricted to ionized regions.

To calculate the number density of ionizing photons $\dot{n}_\gamma$
that escape into the IGM, we first must calculate the fraction $f_{\rm
  c}$ of matter that has collapsed into DM halos and then luminous
objects.  Most of the uncertainties in reionization modeling
originates from the halo-galaxy connection.  More specifically, given
a DM halo, the following main ingredients are needed to calculate the
ionizing emissivity: (i) the star formation rate or stellar mass, (ii)
the stellar ionizing efficiency (i.e. number of ionizing photons per
stellar baryon), and (iii) the UV escape fraction $f_{\rm esc}$.  For
a time and halo mass independent model \citep[e.g.][]{Madau99,
  Robertson15}, $\dot{n}_\gamma = f_{\rm esc} f_\gamma
\bar{\rho}_\star$, where $\bar{\rho}_\star$ is the star formation rate
density and recall that $f_\gamma$ is the photon to stellar baryon
ratio.  However, simulations have shown that star formation rates and
$f_{\rm esc}$ values are strong functions of halo mass $M_{\rm vir}$.
A more accurate value for the ionizing emissivity can be calculated by
integrating over all halo masses,
\begin{equation}
  \dot{n}_\gamma = \int_{M_{\rm min}}^\infty f_{\rm esc} f_\gamma
    (f_{\rm gas} f_\star \dot{f}_{\rm c} M_{\rm vir}) dM_{\rm vir},
\end{equation}
where {\it all} of the factors are functions of halo mass, and the
product inside the parentheses is the star formation rate density in
halo masses between $M$ and $M + dM$.  Here $M_{\rm min}$ is the
minimum halo mass that hosts star formation; $f_{\rm gas} \equiv
M_{\rm gas}/M_{\rm vir}$ is the gas fraction in a halo; $f_\star
\equiv M_\star / M_{\rm gas}$ is the star formation efficiency from
this gas, and $\dot{f}_{\rm c}$ is the time derivative of the
collapsed mass fraction in halos.  This integral can be solved
numerically if given functional forms (either smooth or piecewise
\citep[e.g.][]{Alvarez12, Wise14}), which are usually
computed from simulations \citep[e.g.][]{Chen14} or semi-analytic
models \citep[e.g.][]{Benson06} of early galaxy formation.

\subsection{Semi-numeric models}
\label{sec:semi}

The latest observational constraints strongly suggest that
reionization is driven by galaxies and their UV radiation, and thus it
is a relatively local process.  Semi-numeric models originated from an
analytical (excursion set) treatment of reionization
\citep{Furlanetto04} that made the fundamental assumption that
overdense regions drive reionization.  With this assumption, one
asserts that if the number of ionizing photons, corrected for
recombinations, exceeds the number of baryons in some region, then the
region must be ionized.  This model has been extended to
three-dimensional volumes \citep[e.g.][]{Zahn07, Mesinger07,
  Alvarez12} and can accurately generate full density, velocity, and
ionization fields without the need to follow the underlying physics.

From a single set of cosmological initial conditions, usually given in
a 3-D lattice with $N_{\rm cell}^3$ computational cells, at high
redshift $z \sim 100$, semi-numeric models can compute the ionized
fraction for each cell.  One major advantage to this method is that
these models can calculate $x_{\rm e}(\mathbf{r})$ at any redshift
without performing a time-dependent integral.  Namely, a cell is
considered to be ionized when
\begin{equation}
  f_{\rm c}(\mathrm{r}, M_{\rm min}, R, z) \ge \zeta^{-1},
\end{equation}
where $\zeta$ is the ionization efficiency, and $f_{\rm c}$ is the
collapsed mass fraction inside a sphere of radius $R$ in halos with
masses $M > M_{\rm min}$.  This value of $R$ is iterated from a large
length, usually the UV photon mean free path, down to the width of a
single cell.  The ionization efficiency can be parameterized
\citep[e.g.][]{Greig16}, not unlike analytical models, to be dependent
on the galaxy properties.  This method can fully determine the
ionization state at any time and position in the volume.  The most
computationally intensive portion of semi-numeric calculations
involves Fast Fourier Transforms (FFTs) and usually takes on the order
of core-hours for a single $2048^3$ realization, including the initial
condition generation.

\subsection{Cosmological radiative transfer simulations}

Cosmological $N$-body simulations evolve the cosmic density field and
associated halos, which are then used to calculate the evolution of
time- and space-dependent ionization fields.  They assume that gas
follows the DM, which is a good assumption on large-scales and in
larger halos ($M \gsim 10^9 \Ms$), and then populate these halos with
galaxies that produce ionizing radiation.  It is transported through
the IGM, usually with ray tracing methods that solve the cosmological
radiative transfer equation, which in comoving coordinates
\citep{Gnedin97} is
\begin{equation}
  \label{eqn:rteqn}
  \frac{1}{c} \; \frac{\partial I_\nu}{\partial t} + 
  \frac{\hat{\mathbf{n}} \cdot \nabla I_\nu}{\bar{a}} -
  \frac{H}{c} \; \left( \nu \frac{\partial I_\nu}{\partial \nu} -
  3 I_\nu \right) = -\kappa_\nu I_\nu + j_\nu ,
\end{equation}
reproducing inhomogeneous reionization.  Here $I_\nu \equiv I(\nu,
\mathbf{x}, \Omega, t)$ is the radiation specific intensity in units
of energy per time $t$ per solid angle per unit area per frequency
$\nu$.  $H = \dot{a}/a$ is the Hubble parameter.  $\bar{a} = a/a_{em}$
is the ratio of scale factors at the current time and time of
emission.  The second term represents the propagation of radiation,
where the factor $1/\bar{a}$ accounts for cosmic expansion.  The third
term describes both the cosmological redshift and dilution of
radiation.  On the right hand side, the first term considers the
absorption coefficient $\kappa_\nu \equiv
\kappa_\nu(\mathbf{x},\nu,t)$.  The second term $j_\nu \equiv
j_\nu(\mathbf{x},\nu,t)$ is the emission coefficient that includes any
point sources of radiation or diffuse radiation.

Solving this equation is difficult because of its high dimensionality;
however, we can make some appropriate approximations to reduce its
complexity in order to include radiation transport in numerical
calculations.  Typically timesteps in dynamic calculations are small
enough so that $\Delta a/a \ll 1$, therefore $\bar{a} = 1$ in any
given timestep, reducing the second term to $\hat{\mathbf{n}} \partial
I_\nu/\partial \mathbf{x}$.  To determine the importance of the third
term, we evaluate the ratio of the third term to the second term.
This is $HL/c$, where $L$ is the simulation box length.  If this ratio
is $\ll 1$, we can ignore the third term.  For example at $z=5$, this
ratio is 0.1 when $L = c/H(z=5)$ = 53 proper Mpc.  In large boxes
where the light crossing time is comparable to the Hubble time, then
it becomes important to consider cosmological redshifting and dilution
of the radiation.  Thus equation (\ref{eqn:rteqn}) reduces to the
non-cosmological form in this local approximation,
\begin{equation}
  \label{eqn:localrt}
  \frac{1}{c} \frac{\partial I_\nu}{\partial t} + 
  \hat{\mathbf{n}} \frac{\partial I_\nu}{\partial \mathbf{x}} =
  -\kappa_\nu I_\nu + j_\nu .
\end{equation}
Ray tracing methods represent the source term $j_\nu$ as point sources
of radiation (e.g. stars, galaxies, quasars) that emit radial rays
that are propagated along the direction $\hat{\mathbf{n}}$.

The downside to this method is that it neglects any hydrodynamics and
must make assumptions about the ionizing luminosity escaping from the
halos, the IGM clumping factor $C$, and the suppression of star
formation in low-mass ($M \lsim 10^9 \Ms$) halos.  Such calculations
are performed in either (i) post-processing with the radiative
transfer calculated on density field and halo catalog written to disk,
or (ii) inline where the halo catalogs and radiation sources are
computed on-the-fly, and radiation is traced through the density field
that is stored in memory.  The largest simulations to-date have over
100 billion particles and simulate domains of over 500 comoving Mpc on
a side, and such simulations consume a couple of million core-hours
\citep[e.g.][]{Iliev14}.  They produce similar results as semi-numeric
models but with a larger dynamic range, with higher resolution in
collapsed regions, thus can follow the small-scale ionization
fluctuations to greater accuracy.

\subsection{Full radiation hydrodynamics simulations}

Perhaps the most accurate and computationally expensive calculations
are full radiation hydrodynamics simulations of cosmological galaxy
formation and reionization.  Only in the past decade or so,
computational resources have become large enough, along with
algorithmic advances, to cope with the requirements of such
calculations.  There are two popular methods to solve the radiative
transfer equation coupled to hydrodynamics in three dimensions:
\begin{itemize}
\item {\it Moment methods:} The angular moments of the radiation field
  describe its angular structure, which are related to energy, flux,
  and radiation pressure \citep{Auer70}.  These have been implemented
  in conjunction with short characteristics \citep{Davis12}, with long
  characteristics \citep{Finlator09}, with a variable Eddington tensor
  in the optically-thin limit \citep{Gnedin01} and the general case
  \citep{Jiang14}, and with an M1 closure relation \citep{Rosdahl15,
    Aubert15}.  Moment methods have the advantage of being efficient
  and independent of the number of radiation sources.  However, they
  are diffusive and result in incorrect shadows in some situations.
\item {\it Ray tracing:} Radiation can be propagated along rays that
  extend through a computational grid \citep[e.g.][]{Whalen06,
    Krumholz07, Wise11} or particle set \citep[e.g.][]{Susa06,
    Pawlik08, Hasegawa09}, as discussed previously.  In general, these
  methods are very accurate but computationally expensive because the
  radiation field must be well sampled by the rays with respect to
  the spatial resolution of the domain.
\end{itemize}

In addition to following the DM dynamics, like in the radiative
transfer simulations, they follow the hydrodynamics of the
cosmological domain that allows for the treatment of gaseous collapses
within halos that are driven by radiative cooling.  These radiative
processes are computed through a non-equilibrium chemical network
\citep[e.g.][]{Grackle}.  However computational run-time and memory
limits the resolution of these simulations, which are typically $\sim
10$ times more expensive than the cosmological radiative transfer
simulations.  Depending on the domain size and resolution,
'sub-grid' star formation prescriptions spawn particles that
represent either entire galaxies or individual stellar clusters.
Based on this prescription, the particle has an ionizing luminosity,
whose radiation is the source of the radiative transfer equation
(Equation \ref{eqn:rteqn} or \ref{eqn:localrt}).  Because the
radiation transport is coupled with the hydrodynamics, this equation
must be solved with either small timesteps and/or the appropriate
approximations \citep{Wise11}.  Thus, the radiation sources and the
ensuing hydrodynamic response can be modeled without relying on a
halo-galaxy relationship.  The suppression of star formation,
especially in low-mass galaxies can be directly modeled, along with
the regulation of star formation that results in a more accurate
description of the ionizing sources during the EoR and the process of
reionization itself.  That being said, there still exists
uncertainties, arising from the sub-grid models and convergence issues
of the numerical solvers with respect to resolution.

\section{Conclusions}

Cosmic reionization is the last cosmological phase transition.  There
is strong observational evidence that it ended nearly one billion
years after the Big Bang.  The first generations of galaxies primarily
powered this grand event in the cosmic timeline.  Because structure
forms hierarchically, these first galaxies are the building blocks of
all galaxies we see today, and their properties are passed along as
galaxies assemble.  Thus, further constraints from the epoch of
reionization will play a key role in solidifying theories of galaxy
formation and cosmology.

However there still are unanswered questions in its exact timing, its
progression, its nature, and the role of the first galaxies played
during the epoch of reionization.  These questions will be elucidated
with the upcoming James Webb Space Telescope, ground-based 30-m class
telescopes, and more accurate CMB and 21-cm experiments, all set to be
commissioned within the next decade.  Observing both the reionizing
universe and the galaxies responsible for this transition is paramount
in augmenting our knowledge of this formative period in the universe.

\section*{Acknowledgments}

Many throughout my academic career have contributed to my
understanding of the first stars, first galaxies, radiation transport,
and reionization, but I want to give special thanks to Tom Abel,
Marcelo Alvarez, Renyue Cen, Andrea Ferrara, Andrei Mesinger, Michael
Norman, Brian O'Shea, John Regan, Britton Smith, and Matthew Turk.  My
research is currently supported by National Science Foundation (NSF)
grants AST-1614333 and OAC-1835213, NASA grant NNX17AG23G, and Hubble
theory grant HST-AR-14326.

\section*{Notes on contributor}

\begin{wrapfigure}{l}{0.2\textwidth}
  \centering
  \vspace{-2em}
  \begin{minipage}{0.19\textwidth}
    \includegraphics[width=\textwidth]{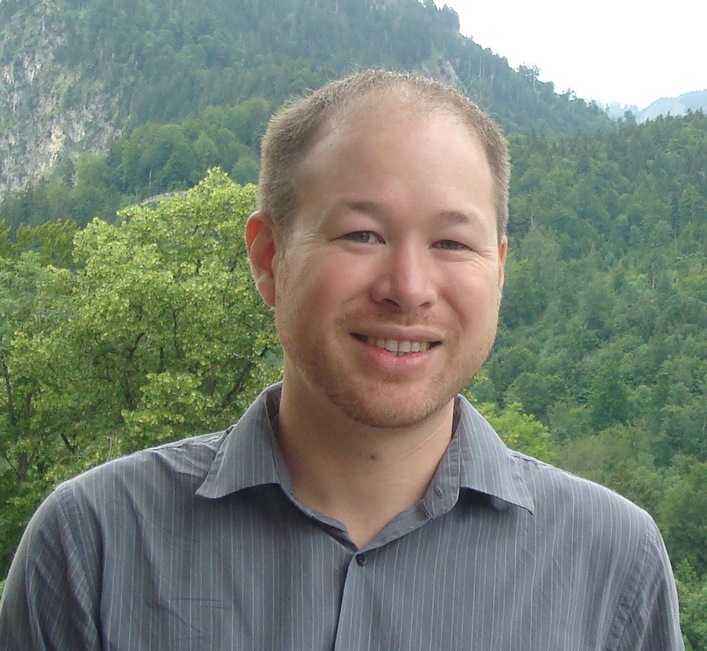}
    \vspace{-1.0cm}
  \end{minipage}
  %\vspace{-6ex}
\end{wrapfigure}

John Wise is the Dunn Family Associate Professor in the School of
Physics and Center for Relativistic Astrophysics at the Georgia
Institute of Technology.  He uses numerical simulations to study the
formation and evolution of galaxies and their black holes.  He is one
of the lead developers of the community-driven, open-source
astrophysics code Enzo
(\href{http://enzo-project.org}{enzo-project.org}).  He received his
B.S. in Physics from the Georgia Tech in 2001. He then studied at
Stanford University, where he received his Ph.D. in Physics in
2007. He went on to work at NASA's Goddard Space Flight Center as a
NASA Postdoctoral Fellow. Then in 2009, he was awarded the Hubble
Fellowship which he took to Princeton University before arriving at
Georgia Tech in 2011, coming back home after ten years roaming the
nation.

\bibliographystyle{tfnlm}
\bibliography{ms}

\begin{thebibliography}{100}
\providecommand{\url}[1]{\normalfont{#1}}
\providecommand{\urlprefix}{Available from: }

\bibitem{Minkowski60}
{Minkowski}~R. {A New Distant Cluster of Galaxies.} \apj. 1960
  Nov;\hspace{0pt}132:908--910.

\bibitem{Matthews61}
{Matthews}~TA, {Bolton}~JG, {Greenstein}~JL, et~al. {paper presented at the
  107th meeting of the AAS}. In: American Astronomical Society Meeting
  Abstracts; Vol. 107; Dec.; 1960.

\bibitem{Schmidt63}
{Schmidt}~M. {3C 273 : A Star-Like Object with Large Red-Shift}. \nat. 1963
  Mar;\hspace{0pt}197:1040.

\bibitem{Hoyle63}
{Hoyle}~F, {Fowler}~WA. {On the nature of strong radio sources}. \mnras.
  1963;\hspace{0pt}125:169.

\bibitem{Schmidt64}
{Schmidt}~M, {Matthews}~TA. {Redshift of the Quasi-Stellar Radio Sources 3c 47
  and 3c 147.} \apj. 1964 Feb;\hspace{0pt}139:781.

\bibitem{Sandage65}
{Sandage}~A. {The Existence of a Major New Constituent of the Universe: the
  Quasistellar Galaxies.} \apj. 1965 May;\hspace{0pt}141:1560.

\bibitem{Osterbrock66}
{Osterbrock}~DE, {Parker}~RAR. {Excitation of the Optical Emission Lines in
  Quasi-Stellar Radio Sources}. \apj. 1966 Jan;\hspace{0pt}143:268.

\bibitem{Salpeter64}
{Salpeter}~EE. {Accretion of Interstellar Matter by Massive Objects.} \apj.
  1964 Aug;\hspace{0pt}140:796--800.

\bibitem{Zeldovich64}
{Zel'dovich}~YB. {The Fate of a Star and the Evolution of Gravitational Energy
  Upon Accretion}. Soviet Physics Doklady. 1964 Sep;\hspace{0pt}9:195.

\bibitem{LB69}
{Lynden-Bell}~D. {Galactic Nuclei as Collapsed Old Quasars}. \nat. 1969
  Aug;\hspace{0pt}223:690--694.

\bibitem{Tananbaum79}
{Tananbaum}~H, {Avni}~Y, {Branduardi}~G, et~al. {X-ray studies of quasars with
  the Einstein Observatory}. \apjl. 1979 Nov;\hspace{0pt}234:L9--L13.

\bibitem{GP65}
{Gunn}~JE, {Peterson}~BA. {On the Density of Neutral Hydrogen in Intergalactic
  Space.} \apj. 1965 Nov;\hspace{0pt}142:1633--1641.

\bibitem{Planck18_Cosmo}
{Planck Collaboration}, {Aghanim}~N, {Akrami}~Y, et~al. {Planck 2018 results.
  VI. Cosmological parameters}. arXiv e-prints. 2018
  Jul;\hspace{0pt}:arXiv:1807.06209.

\bibitem{Rees97}
{Rees}~MJ. {The Universe at z {$>$} 5: When and How Did the 'Dark Age' End?}
  In: {N~R~Tanvir, A~Aragon-Salamanca, \& J~V~Wall}, editor. The Hubble Space
  Telescope and the High Redshift Universe; 1997. p. 115--+.

\bibitem{Skillman14}
{Skillman}~SW, {Warren}~MS, {Turk}~MJ, et~al. {Dark Sky Simulations: Early Data
  Release}. ArXiv e-prints (14072600). 2014 Jul;\hspace{0pt}.

\bibitem{Peebles82}
{Peebles}~PJE. {Large-scale background temperature and mass fluctuations due to
  scale-invariant primeval perturbations}. \apjl. 1982
  Dec;\hspace{0pt}263:L1--L5.

\bibitem{Blumenthal84}
{Blumenthal}~GR, {Faber}~SM, {Primack}~JR, et~al. {Formation of galaxies and
  large-scale structure with cold dark matter}. \nat. 1984
  Oct;\hspace{0pt}311:517--525.

\bibitem{Davis85}
{Davis}~M, {Efstathiou}~G, {Frenk}~CS, et~al. {The evolution of large-scale
  structure in a universe dominated by cold dark matter}. \apj. 1985
  May;\hspace{0pt}292:371--394.

\bibitem{LB67}
{Lynden-Bell}~D. {Statistical mechanics of violent relaxation in stellar
  systems}. \mnras. 1967;\hspace{0pt}136:101--+.

\bibitem{Gunn72}
{Gunn}~JE, {Gott}~JR~III. {On the Infall of Matter Into Clusters of Galaxies
  and Some Effects on Their Evolution}. \apj. 1972 Aug;\hspace{0pt}176:1.

\bibitem{Stroemgren39}
{Str{\"o}mgren}~B. {The Physical State of Interstellar Hydrogen.} \apj. 1939
  May;\hspace{0pt}89:526--+.

\bibitem{Dave01}
{Dav{\'e}}~R, {Cen}~R, {Ostriker}~JP, et~al. {Baryons in the Warm-Hot
  Intergalactic Medium}. \apj. 2001 May;\hspace{0pt}552:473--483.

\bibitem{Haardt12}
{Haardt}~F, {Madau}~P. {Radiative Transfer in a Clumpy Universe. IV. New
  Synthesis Models of the Cosmic UV/X-Ray Background}. \apj. 2012
  Feb;\hspace{0pt}746:125.

\bibitem{Gnedin00}
{Gnedin}~NY. {Effect of Reionization on Structure Formation in the Universe}.
  \apj. 2000 Oct;\hspace{0pt}542:535--541.

\bibitem{Meiksin93}
{Meiksin}~A, {Madau}~P. {On the photoionization of the intergalactic medium by
  quasars at high redshift}. \apj. 1993 Jul;\hspace{0pt}412:34--55.

\bibitem{Tselia10}
{Tseliakhovich}~D, {Hirata}~C. {Relative velocity of dark matter and baryonic
  fluids and the formation of the first structures}. \prd. 2010
  Oct;\hspace{0pt}82(8):083520.

\bibitem{Peebles93}
{Peebles}~PJE. {Principles of Physical Cosmology}. ; 1993.

\bibitem{Barkana01}
{Barkana}~R, {Loeb}~A. {In the beginning: the first sources of light and the
  reionization of the universe}. \physrep. 2001 Jul;\hspace{0pt}349:125--238.

\bibitem{Asplund09}
{Asplund}~M, {Grevesse}~N, {Sauval}~AJ, et~al. {The Chemical Composition of the
  Sun}. Annual Review of Astronomy and Astrophysics. 2009
  Sep;\hspace{0pt}47(1):481--522.

\bibitem{Bromm01}
{Bromm}~V, {Ferrara}~A, {Coppi}~PS, et~al. {The fragmentation of pre-enriched
  primordial objects}. \mnras. 2001 Dec;\hspace{0pt}328:969--976.

\bibitem{ABN02}
{Abel}~T, {Bryan}~GL, {Norman}~ML. {The Formation of the First Star in the
  Universe}. Science. 2002 Jan;\hspace{0pt}295:93--98.

\bibitem{Machacek01}
{Machacek}~ME, {Bryan}~GL, {Abel}~T. {Simulations of Pregalactic Structure
  Formation with Radiative Feedback}. \apj. 2001 Feb;\hspace{0pt}548:509--521.

\bibitem{OLeary12}
{O'Leary}~RM, {McQuinn}~M. {The Formation of the First Cosmic Structures and
  the Physics of the z \~{} 20 Universe}. \apj. 2012 Nov;\hspace{0pt}760:4.

\bibitem{Schauer19}
{Schauer}~ATP, {Glover}~SCO, {Klessen}~RS, et~al. {The influence of streaming
  velocities on the formation of the first stars}. \mnras. 2019
  Apr;\hspace{0pt}484(3):3510--3521.

\bibitem{Wise12}
{Wise}~JH, {Turk}~MJ, {Norman}~ML, et~al. {The Birth of a Galaxy: Primordial
  Metal Enrichment and Stellar Populations}. \apj. 2012 Jan;\hspace{0pt}745:50.

\bibitem{Hirano15}
{Hirano}~S, {Hosokawa}~T, {Yoshida}~N, et~al. {Primordial star formation under
  the influence of far ultraviolet radiation: 1540 cosmological haloes and the
  stellar mass distribution}. \mnras. 2015 Mar;\hspace{0pt}448:568--587.

\bibitem{Turk09}
{Turk}~MJ, {Abel}~T, {O'Shea}~B. {The Formation of Population III Binaries from
  Cosmological Initial Conditions}. Science. 2009 Jul;\hspace{0pt}325:601--.

\bibitem{Greif12}
{Greif}~TH, {Bromm}~V, {Clark}~PC, et~al. {Formation and evolution of
  primordial protostellar systems}. \mnras. 2012 Jul;\hspace{0pt}424:399--415.

\bibitem{Schaerer02}
{Schaerer}~D. {On the properties of massive Population III stars and metal-free
  stellar populations}. \aap. 2002 Jan;\hspace{0pt}382:28--42.

\bibitem{Alvarez06}
{Alvarez}~MA, {Bromm}~V, {Shapiro}~PR. {The H II Region of the First Star}.
  \apj. 2006 Mar;\hspace{0pt}639:621--632.

\bibitem{Kimm14}
{Kimm}~T, {Cen}~R. {Escape Fraction of Ionizing Photons during Reionization:
  Effects due to Supernova Feedback and Runaway OB Stars}. \apj. 2014
  Jun;\hspace{0pt}788:121.

\bibitem{Kimm16}
{Kimm}~T, {Katz}~H, {Haehnelt}~M, et~al. {Feedback-regulated star formation and
  escape of LyC photons from mini-haloes during reionization}. \mnras. 2017
  Apr;\hspace{0pt}466(4):4826--4846.

\bibitem{Xu16}
{Xu}~H, {Wise}~JH, {Norman}~ML, et~al. {Galaxy Properties and UV Escape
  Fractions during the Epoch of Reionization: Results from the Renaissance
  Simulations}. \apj. 2016 Dec;\hspace{0pt}833:84.

\bibitem{Wise14}
{Wise}~JH, {Demchenko}~VG, {Halicek}~MT, et~al. {The birth of a galaxy - III.
  Propelling reionization with the faintest galaxies}. \mnras. 2014
  Aug;\hspace{0pt}442:2560--2579.

\bibitem{Robertson13}
{Robertson}~BE, {Furlanetto}~SR, {Schneider}~E, et~al. {New Constraints on
  Cosmic Reionization from the 2012 Hubble Ultra Deep Field Campaign}. \apj.
  2013 May;\hspace{0pt}768:71.

\bibitem{Ma15}
{Ma}~X, {Kasen}~D, {Hopkins}~PF, et~al. {The difficulty of getting high escape
  fractions of ionizing photons from high-redshift galaxies: a view from the
  FIRE cosmological simulations}. \mnras. 2015 Oct;\hspace{0pt}453:960--975.

\bibitem{Xu14}
{Xu}~H, {Ahn}~K, {Wise}~JH, et~al. {Heating the Intergalactic Medium by X-Rays
  from Population III Binaries in High-redshift Galaxies}. \apj. 2014
  Aug;\hspace{0pt}791:110.

\bibitem{Mesinger13}
{Mesinger}~A, {Ferrara}~A, {Spiegel}~DS. {Signatures of X-rays in the early
  Universe}. \mnras. 2013 May;\hspace{0pt}431:621--637.

\bibitem{Banados18}
{Ba{\~n}ados}~E, {Venemans}~BP, {Mazzucchelli}~C, et~al. {An
  800-million-solar-mass black hole in a significantly neutral Universe at a
  redshift of 7.5}. \nat. 2018 Jan;\hspace{0pt}553:473--476.

\bibitem{Fan06b}
{Fan}~X, {Strauss}~MA, {Becker}~RH, et~al. {Constraining the Evolution of the
  Ionizing Background and the Epoch of Reionization with z\~{}6 Quasars. II. A
  Sample of 19 Quasars}. \aj. 2006 Jul;\hspace{0pt}132:117--136.

\bibitem{Alvarez09}
{Alvarez}~MA, {Wise}~JH, {Abel}~T. {Accretion onto the First Stellar-Mass Black
  Holes}. \apjl. 2009 Aug;\hspace{0pt}701:L133--L137.

\bibitem{Greig16_QSOz7}
{Greig}~B, {Mesinger}~A, {Haiman}~Z, et~al. {Are we witnessing the epoch of
  reionisation at z = 7.1 from the spectrum of J1120+0641?} \mnras. 2017
  Apr;\hspace{0pt}466(4):4239--4249.

\bibitem{Becker15}
{Becker}~GD, {Bolton}~JS, {Madau}~P, et~al. {Evidence of patchy hydrogen
  reionization from an extreme Ly{$\alpha$} trough below redshift six}. \mnras.
  2015 Mar;\hspace{0pt}447:3402--3419.

\bibitem{Songaila04}
{Songaila}~A. {The Evolution of the Intergalactic Medium Transmission to
  Redshift 6}. \aj. 2004 May;\hspace{0pt}127:2598--2603.

\bibitem{Mesinger10}
{Mesinger}~A. {Was reionization complete by z \~{} 5-6?} \mnras. 2010
  Sep;\hspace{0pt}407:1328--1337.

\bibitem{Greig16}
{Greig}~B, {Mesinger}~A. {The global history of reionization}. \mnras. 2017
  Mar;\hspace{0pt}465(4):4838--4852.

\bibitem{McDonald01}
{McDonald}~P, {Miralda-Escud{\'e}}~J, {Rauch}~M, et~al. {A Measurement of the
  Temperature-Density Relation in the Intergalactic Medium Using a New
  Ly{$\alpha$} Absorption-Line Fitting Method}. \apj. 2001
  Nov;\hspace{0pt}562:52--75.

\bibitem{York00}
{York}~DG, {Adelman}~J, {Anderson}~JE~Jr, et~al. {The Sloan Digital Sky Survey:
  Technical Summary}. \aj. 2000 Sep;\hspace{0pt}120:1579--1587.

\bibitem{Becker13}
{Becker}~GD, {Bolton}~JS. {New measurements of the ionizing ultraviolet
  background over 2 {$<$} z {$<$} 5 and implications for hydrogen
  reionization}. \mnras. 2013 Dec;\hspace{0pt}436:1023--1039.

\bibitem{Schaye00}
{Schaye}~J, {Theuns}~T, {Rauch}~M, et~al. {The thermal history of the
  intergalactic medium$^{*}$}. \mnras. 2000 Nov;\hspace{0pt}318:817--826.

\bibitem{Hui97}
{Hui}~L, {Gnedin}~NY. {Equation of state of the photoionized intergalactic
  medium}. \mnras. 1997 Nov;\hspace{0pt}292:27.

\bibitem{Bolton10}
{Bolton}~JS, {Becker}~GD, {Wyithe}~JSB, et~al. {A first direct measurement of
  the intergalactic medium temperature around a quasar at z = 6}. \mnras. 2010
  Jul;\hspace{0pt}406:612--625.

\bibitem{Bolton12}
{Bolton}~JS, {Becker}~GD, {Raskutti}~S, et~al. {Improved measurements of the
  intergalactic medium temperature around quasars: possible evidence for the
  initial stages of He II reionization at z {$\simeq$} 6}. \mnras. 2012
  Feb;\hspace{0pt}419:2880--2892.

\bibitem{Ewen51}
{Ewen}~HI, {Purcell}~EM. {Observation of a Line in the Galactic Radio Spectrum:
  Radiation from Galactic Hydrogen at 1,420 Mc./sec.} \nat. 1951
  Sep;\hspace{0pt}168:356.

\bibitem{Muller51}
{Muller}~CA, {Oort}~JH. {Observation of a Line in the Galactic Radio Spectrum:
  The Interstellar Hydrogen Line at 1,420 Mc./sec., and an Estimate of Galactic
  Rotation}. \nat. 1951 Sep;\hspace{0pt}168:357--358.

\bibitem{Scott90}
{Scott}~D, {Rees}~MJ. {The 21-cm line at high redshift: a diagnostic for the
  origin of large scale structure}. \mnras. 1990 Dec;\hspace{0pt}247:510.

\bibitem{Bowman18_EDGES}
{Bowman}~JD, {Rogers}~AEE, {Monsalve}~RA, et~al. {An absorption profile centred
  at 78 megahertz in the sky-averaged spectrum}. \nat. 2018
  Mar;\hspace{0pt}555(7694):67--70.

\bibitem{Mirocha19}
{Mirocha}~J, {Furlanetto}~SR. {What does the first highly redshifted 21-cm
  detection tell us about early galaxies?} \mnras. 2019
  Feb;\hspace{0pt}483(2):1980--1992.

\bibitem{Cohen17}
{Cohen}~A, {Fialkov}~A, {Barkana}~R, et~al. {Charting the parameter space of
  the global 21-cm signal}. \mnras. 2017 Dec;\hspace{0pt}472(2):1915--1931.

\bibitem{Kaurov18}
{Kaurov}~AA, {Venumadhav}~T, {Dai}~L, et~al. {Implication of the Shape of the
  EDGES Signal for the 21 cm Power Spectrum}. \apj. 2018
  Sep;\hspace{0pt}864(1):L15.

\bibitem{Barkana18}
{Barkana}~R. {Possible interaction between baryons and dark-matter particles
  revealed by the first stars}. \nat. 2018 Mar;\hspace{0pt}555(7694):71--74.

\bibitem{paper}
{Parsons}~AR, {Backer}~DC, {Foster}~GS, et~al. {The Precision Array for Probing
  the Epoch of Re-ionization: Eight Station Results}. \aj. 2010
  Apr;\hspace{0pt}139:1468--1480.

\bibitem{lofar}
{van Haarlem}~MP, {Wise}~MW, {Gunst}~AW, et~al. {LOFAR: The LOw-Frequency
  ARray}. \aap. 2013 Aug;\hspace{0pt}556:A2.

\bibitem{mwa}
{Bowman}~JD, {Cairns}~I, {Kaplan}~DL, et~al. {Science with the Murchison
  Widefield Array}. \pasa. 2013 Apr;\hspace{0pt}30:e031.

\bibitem{hera}
{Neben}~AR, {Bradley}~RF, {Hewitt}~JN, et~al. {The Hydrogen Epoch of
  Reionization Array Dish. I. Beam Pattern Measurements and Science
  Implications}. \apj. 2016 Aug;\hspace{0pt}826:199.

\bibitem{ska}
Dewdney~PE, Hall~PJ, Schilizzi~RT, et~al. The square kilometre array.
  Proceedings of the IEEE. 2009;\hspace{0pt}97(8):1482--1496.

\bibitem{Kashikawa15}
{Kashikawa}~N, {Ishizaki}~Y, {Willott}~CJ, et~al. {The Subaru High-z Quasar
  Survey: Discovery of Faint z \~{} 6 Quasars}. \apj. 2015
  Jan;\hspace{0pt}798:28.

\bibitem{Willott10}
{Willott}~CJ, {Albert}~L, {Arzoumanian}~D, et~al. {Eddington-limited Accretion
  and the Black Hole Mass Function at Redshift 6}. \aj. 2010
  Aug;\hspace{0pt}140:546--560.

\bibitem{Grissom14}
{Grissom}~RL, {Ballantyne}~DR, {Wise}~JH. {On the contribution of active
  galactic nuclei to reionization}. \aap. 2014 Jan;\hspace{0pt}561:A90.

\bibitem{Madau15}
{Madau}~P, {Haardt}~F. {Cosmic Reionization after Planck: Could Quasars Do It
  All?} \apjl. 2015 Nov;\hspace{0pt}813:L8.

\bibitem{Ellis13}
{Ellis}~RS, {McLure}~RJ, {Dunlop}~JS, et~al. {The Abundance of Star-forming
  Galaxies in the Redshift Range 8.5-12: New Results from the 2012 Hubble Ultra
  Deep Field Campaign}. \apjl. 2013 Jan;\hspace{0pt}763:L7.

\bibitem{Coe15}
{Coe}~D, {Bradley}~L, {Zitrin}~A. {Frontier Fields: High-redshift Predictions
  and Early Results}. \apj. 2015 Feb;\hspace{0pt}800:84.

\bibitem{Laporte16}
{Laporte}~N, {Infante}~L, {Troncoso Iribarren}~P, et~al. {Young Galaxy
  Candidates in the Hubble Frontier Fields. III. MACS J0717.5+3745}. \apj. 2016
  Apr;\hspace{0pt}820:98.

\bibitem{Oesch16}
{Oesch}~PA, {Brammer}~G, {van Dokkum}~PG, et~al. {A Remarkably Luminous Galaxy
  at z=11.1 Measured with Hubble Space Telescope Grism Spectroscopy}. \apj.
  2016 Mar;\hspace{0pt}819:129.

\bibitem{Schechter76}
{Schechter}~P. {An analytic expression for the luminosity function for
  galaxies.} \apj. 1976 Jan;\hspace{0pt}203:297--306.

\bibitem{McLure13}
{McLure}~RJ, {Dunlop}~JS, {Bowler}~RAA, et~al. {A new multifield determination
  of the galaxy luminosity function at z = 7-9 incorporating the 2012 Hubble
  Ultra-Deep Field imaging}. \mnras. 2013 Jul;\hspace{0pt}432:2696--2716.

\bibitem{Livermore16}
{Livermore}~RC, {Finkelstein}~SL, {Lotz}~JM. {Directly Observing the Galaxies
  Likely Responsible for Reionization}. \apj. 2017 Feb;\hspace{0pt}835(2):113.

\bibitem{Bouwens16}
{Bouwens}~RJ, {Oesch}~PA, {Illingworth}~GD, et~al. {The z ̃ 6 Luminosity
  Function Fainter than -15 mag from the Hubble Frontier Fields: The Impact of
  Magnification Uncertainties}. \apj. 2017 Jul;\hspace{0pt}843(2):129.

\bibitem{Finkelstein12}
{Finkelstein}~SL, {Papovich}~C, {Ryan}~RE, et~al. {CANDELS: The Contribution of
  the Observed Galaxy Population to Cosmic Reionization}. \apj. 2012
  Oct;\hspace{0pt}758:93.

\bibitem{Siana15}
{Siana}~B, {Shapley}~AE, {Kulas}~KR, et~al. {A Deep Hubble Space Telescope and
  Keck Search for Definitive Identification of Lyman Continuum Emitters at
  z\~{}3.1}. \apj. 2015 May;\hspace{0pt}804:17.

\bibitem{Nestor13}
{Nestor}~DB, {Shapley}~AE, {Kornei}~KA, et~al. {A Refined Estimate of the
  Ionizing Emissivity from Galaxies at z \~{}= 3: Spectroscopic Follow-up in
  the SSA22a Field}. \apj. 2013 Mar;\hspace{0pt}765:47.

\bibitem{Cooke14}
{Cooke}~J, {Ryan-Weber}~EV, {Garel}~T, et~al. {Lyman-continuum galaxies and the
  escape fraction of Lyman-break galaxies}. \mnras. 2014
  Jun;\hspace{0pt}441:837--851.

\bibitem{Arons70}
{Arons}~J, {McCray}~R. {Photo-Ionization of Intergalactic Hydrogen by Quasars}.
  \aplett. 1970;\hspace{0pt}5:123.

\bibitem{Pawlik09}
{Pawlik}~AH, {Schaye}~J, {van Scherpenzeel}~E. {Keeping the Universe ionized:
  photoheating and the clumping factor of the high-redshift intergalactic
  medium}. \mnras. 2009 Apr;\hspace{0pt}394:1812--1824.

\bibitem{So14}
{So}~GC, {Norman}~ML, {Reynolds}~DR, et~al. {Fully Coupled Simulation of Cosmic
  Reionization. II. Recombinations, Clumping Factors, and the Photon Budget for
  Reionization}. \apj. 2014 Jul;\hspace{0pt}789:149.

\bibitem{Finlator12}
{Finlator}~K, {Oh}~SP, {{\"O}zel}~F, et~al. {Gas clumping in self-consistent
  reionization models}. \mnras. 2012 Dec;\hspace{0pt}427:2464--2479.

\bibitem{Madau99}
{Madau}~P, {Haardt}~F, {Rees}~MJ. {Radiative Transfer in a Clumpy Universe.
  III. The Nature of Cosmological Ionizing Sources}. \apj. 1999
  Apr;\hspace{0pt}514:648--659.

\bibitem{Robertson15}
{Robertson}~BE, {Ellis}~RS, {Furlanetto}~SR, et~al. {Cosmic Reionization and
  Early Star-forming Galaxies: A Joint Analysis of New Constraints from Planck
  and the Hubble Space Telescope}. \apjl. 2015 Apr;\hspace{0pt}802:L19.

\bibitem{Alvarez12}
{Alvarez}~MA, {Finlator}~K, {Trenti}~M. {Constraints on the Ionizing Efficiency
  of the First Galaxies}. \apjl. 2012 Nov;\hspace{0pt}759:L38.

\bibitem{Chen14}
{Chen}~P, {Wise}~JH, {Norman}~ML, et~al. {Scaling Relations for Galaxies Prior
  to Reionization}. \apj. 2014 Nov;\hspace{0pt}795:144.

\bibitem{Benson06}
{Benson}~AJ, {Sugiyama}~N, {Nusser}~A, et~al. {The epoch of reionization}.
  \mnras. 2006 Jul;\hspace{0pt}369:1055--1080.

\bibitem{Furlanetto04}
{Furlanetto}~SR, {Zaldarriaga}~M, {Hernquist}~L. {The Growth of H II Regions
  During Reionization}. \apj. 2004 Sep;\hspace{0pt}613:1--15.

\bibitem{Zahn07}
{Zahn}~O, {Lidz}~A, {McQuinn}~M, et~al. {Simulations and Analytic Calculations
  of Bubble Growth during Hydrogen Reionization}. \apj. 2007
  Jan;\hspace{0pt}654:12--26.

\bibitem{Mesinger07}
{Mesinger}~A, {Furlanetto}~S. {Efficient Simulations of Early Structure
  Formation and Reionization}. \apj. 2007 Nov;\hspace{0pt}669:663--675.

\bibitem{Gnedin97}
{Gnedin}~NY, {Ostriker}~JP. {Reionization of the Universe and the Early
  Production of Metals}. \apj. 1997 Sep;\hspace{0pt}486:581--+.

\bibitem{Iliev14}
{Iliev}~IT, {Mellema}~G, {Ahn}~K, et~al. {Simulating cosmic reionization: how
  large a volume is large enough?} \mnras. 2014 Mar;\hspace{0pt}439:725--743.

\bibitem{Auer70}
{Auer}~LH, {Mihalas}~D. {On the use of variable Eddington factors in non-LTE
  stellar atmospheres computations}. \mnras. 1970;\hspace{0pt}149:65--+.

\bibitem{Davis12}
{Davis}~SW, {Stone}~JM, {Jiang}~YF. {A Radiation Transfer Solver for Athena
  Using Short Characteristics}. \apjs. 2012 Mar;\hspace{0pt}199:9.

\bibitem{Finlator09}
{Finlator}~K, {{\"O}zel}~F, {Dav{\'e}}~R. {A new moment method for continuum
  radiative transfer in cosmological re-ionization}. \mnras. 2009
  Mar;\hspace{0pt}393:1090--1106.

\bibitem{Gnedin01}
{Gnedin}~NY, {Abel}~T. {Multi-dimensional cosmological radiative transfer with
  a Variable Eddington Tensor formalism}. \na. 2001 Oct;\hspace{0pt}6:437--455.

\bibitem{Jiang14}
{Jiang}~YF, {Stone}~JM, {Davis}~SW. {An Algorithm for Radiation
  Magnetohydrodynamics Based on Solving the Time-dependent Transfer Equation}.
  \apjs. 2014 Jul;\hspace{0pt}213:7.

\bibitem{Rosdahl15}
{Rosdahl}~J, {Teyssier}~R. {A scheme for radiation pressure and photon
  diffusion with the M1 closure in RAMSES-RT}. \mnras. 2015
  Jun;\hspace{0pt}449:4380--4403.

\bibitem{Aubert15}
{Aubert}~D, {Deparis}~N, {Ocvirk}~P. {EMMA: an adaptive mesh refinement
  cosmological simulation code with radiative transfer}. \mnras. 2015
  Nov;\hspace{0pt}454:1012--1037.

\bibitem{Whalen06}
{Whalen}~D, {Norman}~ML. {A Multistep Algorithm for the Radiation
  Hydrodynamical Transport of Cosmological Ionization Fronts and Ionized
  Flows}. \apjs. 2006 Feb;\hspace{0pt}162:281--303.

\bibitem{Krumholz07}
{Krumholz}~MR, {Klein}~RI, {McKee}~CF. {Radiation-Hydrodynamic Simulations of
  Collapse and Fragmentation in Massive Protostellar Cores}. \apj. 2007
  Feb;\hspace{0pt}656:959--979.

\bibitem{Wise11}
{Wise}~JH, {Abel}~T. {ENZO+MORAY: radiation hydrodynamics adaptive mesh
  refinement simulations with adaptive ray tracing}. \mnras. 2011
  Jul;\hspace{0pt}414:3458--3491.

\bibitem{Susa06}
{Susa}~H. {Smoothed Particle Hydrodynamics Coupled with Radiation Transfer}.
  \pasj. 2006 Apr;\hspace{0pt}58:445--460.

\bibitem{Pawlik08}
{Pawlik}~AH, {Schaye}~J. {TRAPHIC - radiative transfer for smoothed particle
  hydrodynamics simulations}. \mnras. 2008 Sep;\hspace{0pt}389:651--677.

\bibitem{Hasegawa09}
{Hasegawa}~K, {Umemura}~M, {Susa}~H. {Radiative regulation of Population III
  star formation}. \mnras. 2009 May;\hspace{0pt}395:1280--1286.

\bibitem{Grackle}
{Smith}~BD, {Bryan}~GL, {Glover}~SCO, et~al. {GRACKLE: a chemistry and cooling
  library for astrophysics}. \mnras. 2017 Apr;\hspace{0pt}466(2):2217--2234.

\end{thebibliography}

\end{document}